\PassOptionsToPackage{table}{xcolor}
\documentclass[sigconf, nonacm]{acmart}
\usepackage{xspace}
\usepackage{booktabs}
\usepackage{tabularx}
\usepackage{multirow}
\usepackage{graphicx}
\usepackage{amsmath}
\usepackage{xcolor}
\usepackage{microtype}
\usepackage{enumitem}
\usepackage{url}
\usepackage{algorithm}
\usepackage{algpseudocode}
\usepackage[most]{tcolorbox}
\newtcolorbox{takeaway}{enhanced,colback=blue!4!white,colframe=blue!55!black,boxrule=0.6pt,arc=2pt,left=4pt,right=4pt,top=1pt,bottom=1pt,toptitle=0.5pt,bottomtitle=0.5pt,before skip=3pt,after skip=3pt,fonttitle=\bfseries\small,title=Key takeaway}
\setlist[itemize]{leftmargin=1.2em,itemsep=1pt,topsep=2pt,parsep=0pt}
\setlist[enumerate]{leftmargin=1.4em,itemsep=1pt,topsep=2pt,parsep=0pt}

\newcommand{\llmRate}{48\xspace}
\newcommand{\nAcc}{19{,}517\xspace}

\newcommand{\nAsevenFound}{149\xspace}

\newcommand{\nAsevenPool}{1{,}933\xspace}

\newcommand{\nAwayHome}{2{,}979\xspace}
\newcommand{\nCCode}{3{,}286\xspace}

\newcommand{\nDupSame}{3{,}466\xspace}
\newcommand{\nDupShared}{849\xspace}
\newcommand{\nDupSurname}{275\xspace}
\newcommand{\nEmail}{17{,}712\xspace}

\newcommand{\nHosts}{9{,}234\xspace}
\newcommand{\nInacc}{8{,}306\xspace}
\newcommand{\nInstAcc}{680\xspace}
\newcommand{\nInstAll}{702\xspace}

\newcommand{\nInstBig}{561\xspace}

\newcommand{\nNoEmailAcc}{1{,}805\xspace}

\newcommand{\nRescueIn}{3{,}573\xspace}
\newcommand{\nRobotsRendered}{50\xspace}
\newcommand{\nSBlocked}{1{,}741\xspace}
\newcommand{\nSHttpError}{227\xspace}
\newcommand{\nSNonHtml}{12\xspace}
\newcommand{\nSNotFound}{2{,}839\xspace}
\newcommand{\nSRobotsDisallowed}{320\xspace}
\newcommand{\nSServerError}{144\xspace}
\newcommand{\nSUnreachable}{3{,}023\xspace}

\newcommand{\nSharedUnique}{1{,}124\xspace}
\newcommand{\nSrcLlm}{2{,}593\xspace}
\newcommand{\nSrcRescue}{1{,}788\xspace}
\newcommand{\nSrcRule}{13{,}331\xspace}

\newcommand{\nTotal}{32{,}413\xspace}
\newcommand{\nUnDns}{794\xspace}
\newcommand{\nUnOther}{382\xspace}
\newcommand{\nUnTimeout}{1{,}038\xspace}
\newcommand{\nUnTls}{809\xspace}
\newcommand{\nUnique}{28{,}947\xspace}
\newcommand{\pAcc}{67.4\%\xspace}

\newcommand{\pAwayHome}{16.8\%\xspace}
\newcommand{\pCCode}{18.6\%\xspace}
\newcommand{\pCFCodeAction}{51.3\%\xspace}

\newcommand{\pCLCodeHomepage}{86.7\%\xspace}

\newcommand{\pCLSemanticContactpage}{13.0\%\xspace}
\newcommand{\pCLSemanticHomepage}{69.6\%\xspace}

\newcommand{\pCPlain}{53.8\%\xspace}
\newcommand{\pCSemantic}{4.1\%\xspace}
\newcommand{\pCSymbolic}{23.5\%\xspace}
\newcommand{\pCodeRender}{5.5\%\xspace}
\newcommand{\pConfLow}{1.0\%\xspace}
\newcommand{\pDupSame}{10.7\%\xspace}

\newcommand{\pEmailAcc}{90.8\%\xspace}
\newcommand{\pFAction}{9.8\%\xspace}
\newcommand{\pFImage}{2.1\%\xspace}
\newcommand{\pFText}{88.2\%\xspace}
\newcommand{\pFallback}{4.6\%\xspace}
\newcommand{\pInacc}{28.7\%\xspace}

\newcommand{\pLCVpage}{0.2\%\xspace}
\newcommand{\pLContactpage}{3.7\%\xspace}
\newcommand{\pLHomepage}{83.2\%\xspace}
\newcommand{\pLOthersamesitepage}{7.7\%\xspace}
\newcommand{\pLOthersite}{1.7\%\xspace}
\newcommand{\pLPDF}{3.6\%\xspace}
\newcommand{\pLadJs}{93.8\%\xspace}
\newcommand{\pLadLex}{75.3\%\xspace}

\newcommand{\pLadRegex}{52.1\%\xspace}
\newcommand{\pLadVision}{95.9\%\xspace}

\newcommand{\pNeedLlm}{4.1\%\xspace}

\newcommand{\pNoEmailAcc}{9.2\%\xspace}

\newcommand{\pSharedUnique}{3.9\%\xspace}
\newcommand{\pSrcLlm}{14.6\%\xspace}

\newcommand{\pSrcRescue}{10.1\%\xspace}
\newcommand{\pSrcRule}{75.3\%\xspace}
\newcommand{\pSubCloudflare}{4.1\%\xspace}
\newcommand{\pSubEncoded}{13.3\%\xspace}
\newcommand{\pSubEntity}{5.5\%\xspace}
\newcommand{\pSubMailto}{38.6\%\xspace}
\newcommand{\pSubRender}{29.7\%\xspace}
\newcommand{\pSubSourceOnly}{6.2\%\xspace}

\IfFileExists{numbers_extra.tex}{

\newcommand{\PSubSemDescription}{9.3\%\xspace}
\newcommand{\PSubSemInstruction}{5.2\%\xspace}

\newcommand{\PSubSemTemplate}{46.0\%\xspace}
\newcommand{\PSubSemWordSubstitution}{15.1\%\xspace}

\newcommand{\VCF}{0.47\xspace}
\newcommand{\VCFCI}{[0.47, 0.48]\xspace}
\newcommand{\VCL}{0.12\xspace}

\newcommand{\VFL}{0.07\xspace}

\newcommand{\nLeak}{509\xspace}

\newcommand{\nNonPlain}{8,186\xspace}

\newcommand{\pLeak}{6.2\%\xspace}

\newcommand{\pSemNonEn}{16.7\%\xspace}

\newcommand{\pThreePlusAddr}{0.6\%\xspace}
\newcommand{\pTopThreeTriples}{70.4\%\xspace}
\newcommand{\pTopTriple}{42.2\%\xspace}
\newcommand{\pTwoAddr}{6.2\%\xspace}

\newcommand{\pairExp}{38.1\%\xspace}
\newcommand{\pairSameHost}{67.9\%\xspace}
\newcommand{\pairSameHostCI}{[62.9, 72.9]\xspace}

\newcommand{\pairSameInstDiffHost}{38.8\%\xspace}
\newcommand{\pairSameInstDiffHostCI}{[37.1, 40.8]\xspace}

\newcommand{\psymBr}{33.4\%\xspace}
\newcommand{\psymChar}{0.5\%\xspace}
\newcommand{\psymOther}{24.3\%\xspace}

\newcommand{\psymWords}{37.4\%\xspace}

\newcommand{\semCheckN}{100\xspace}
\newcommand{\semCheckRate}{88\%\xspace}

}{}
\IfFileExists{numbers_extra2.tex}{

\newcommand{\sigNCf}{547\xspace}
\newcommand{\sigInstCf}{112\xspace}
\newcommand{\sigNPCf}{97\%\xspace}

\newcommand{\sigNPlugin}{172\xspace}

\newcommand{\sigNPPlugin}{91\%\xspace}
}{}
\IfFileExists{numbers_deep.tex}{

\newcommand{\chiCountryP}{<0.001\xspace}

\newcommand{\iccHost}{0.41\xspace}
\newcommand{\iccInstDeep}{0.19\xspace}

\newcommand{\nHostsGEFour}{734\xspace}
\newcommand{\nHostsICC}{1,179\xspace}

\newcommand{\naJp}{20.4\%\xspace}

\newcommand{\naNl}{3.5\%\xspace}

\newcommand{\npCa}{42.4\%\xspace}
\newcommand{\npCn}{37.4\%\xspace}
\newcommand{\npDe}{60.1\%\xspace}

\newcommand{\npFr}{62.1\%\xspace}
\newcommand{\npGb}{39.6\%\xspace}

\newcommand{\npJp}{70.7\%\xspace}

\newcommand{\pHostUniform}{34.2\%\xspace}
\newcommand{\pHostUniformNull}{13.0\%\xspace}

\newcommand{\pRInstFE}{0.180\xspace}
\newcommand{\pRMFour}{0.052\xspace}
\newcommand{\pRMOne}{0.005\xspace}

\newcommand{\swResearchPortalCode}{70\%\xspace}
\newcommand{\swResearchPortalN}{470\xspace}
\newcommand{\swResearchPortalNP}{76\%\xspace}

\newcommand{\swSiteBuilderNP}{52\%\xspace}

\newcommand{\swUniversityCmsNP}{30\%\xspace}

\newcommand{\swWordpressNP}{43\%\xspace}
}{}
\IfFileExists{numbers_sampling.tex}{

}{}
\IfFileExists{numbers_human.tex}{
\newcommand{\hvAgreeContent}{98.4\%\xspace}

\newcommand{\hvAgreeHas}{99.3\%\xspace}

\newcommand{\hvContent}{90.3\%\xspace}

\newcommand{\hvContentN}{124\xspace}

\newcommand{\hvEmailStrict}{94.4\%\xspace}
\newcommand{\hvEmailStrictCI}{[91.2, 96.5]\xspace}
\newcommand{\hvEmailStrictRaw}{92.3\%\xspace}

\newcommand{\hvExtraN}{5\xspace}
\newcommand{\hvForm}{95.2\%\xspace}

\newcommand{\hvKappaContent}{0.98\xspace}

\newcommand{\hvKappaHas}{0.97\xspace}

\newcommand{\hvLocation}{95.2\%\xspace}

\newcommand{\hvMissN}{6\xspace}
\newcommand{\hvN}{150\xspace}
\newcommand{\hvNJudged}{143\xspace}

\newcommand{\hvStrictGoodN}{132\xspace}

}{}
\IfFileExists{numbers_landing2.tex}{
\newcommand{\nClickLand}{2,979\xspace}

\newcommand{\pClickLand}{16.8\%\xspace}

\newcommand{\pClickPdf}{21.3\%\xspace}

\newcommand{\pClickSameHost}{44.9\%\xspace}

}{}
\IfFileExists{numbers_prot.tex}{
\newcommand{\nBlockedHosts}{538\xspace}

\newcommand{\nChalAws}{113\xspace}

\newcommand{\nChalFastly}{267\xspace}

\newcommand{\nDoorChal}{526\xspace}

\newcommand{\nLandNonPersonal}{1,828\xspace}

\newcommand{\nLockAkamai}{24\xspace}
\newcommand{\nLockAnubis}{44\xspace}
\newcommand{\nLockAws}{135\xspace}
\newcommand{\nLockCloudflare}{1,007\xspace}

\newcommand{\nLockImperva}{40\xspace}
\newcommand{\nLockedBrowser}{122\xspace}

\newcommand{\nLockedRefused}{1,741\xspace}
\newcommand{\nRedOtherDom}{1,045\xspace}
\newcommand{\nRedSameDom}{2,188\xspace}
\newcommand{\nRedSameHost}{1,833\xspace}
\newcommand{\nRedirect}{5,066\xspace}

\newcommand{\pDoorChal}{2.5\%\xspace}
\newcommand{\pDoorLocked}{8.8\%\xspace}
\newcommand{\pDoorOpen}{88.8\%\xspace}

\newcommand{\pLandDirectory}{8.6\%\xspace}
\newcommand{\pLandError}{3.8\%\xspace}
\newcommand{\pLandFront}{14.1\%\xspace}
\newcommand{\pLandNonPersonal}{36.1\%\xspace}
\newcommand{\pLandOther}{9.5\%\xspace}

\newcommand{\pLockedIdent}{67.4\%\xspace}

\newcommand{\pRedirect}{26.0\%\xspace}
\newcommand{\pVaultOpenGivenChal}{76.1\%\xspace}
\newcommand{\pVaultOpenGivenOpen}{51.4\%\xspace}

}{}
\IfFileExists{numbers_ic3.tex}{

\newcommand{\combCXFXLEighty}{5\xspace}
\newcommand{\combCXFXLObs}{47\xspace}

\newcommand{\icKclusters}{4\xspace}

\newcommand{\icMailConc}{76\%\xspace}

\newcommand{\icMailInst}{9\xspace}

\newcommand{\icMailTop}{85\%\xspace}

\newcommand{\icMixInst}{177\xspace}

\newcommand{\icMixPlain}{43\%\xspace}

\newcommand{\icMixSch}{7,045\xspace}

\newcommand{\icMixSym}{31\%\xspace}

\newcommand{\icPlainInst}{188\xspace}

\newcommand{\icPlainPlain}{69\%\xspace}

\newcommand{\icRenConc}{80\%\xspace}

\newcommand{\icRenInst}{12\xspace}

\newcommand{\icRenTop}{80\%\xspace}
\newcommand{\icSil}{0.32\xspace}
\newcommand{\nBarrierFixed}{460\xspace}
\newcommand{\nBarrierInst}{71\xspace}
\newcommand{\nInstFifteen}{386\xspace}

\newcommand{\pBarrierTopThree}{21\%\xspace}
}{}
\newcommand{\cPlain}{\textsc{Plain}\xspace}
\newcommand{\cSubst}{\textsc{Symbolic}\xspace}
\newcommand{\cCode}{\textsc{Code}\xspace}
\newcommand{\cSem}{\textsc{Semantic}\xspace}
\newcommand{\fText}{\textsc{Text}\xspace}
\newcommand{\fImage}{\textsc{Image}\xspace}
\newcommand{\fAct}{\textsc{Interaction}\xspace}
\newcommand{\sys}{\textsc{MailScope}\xspace}

\newcommand{\rescueInstPages}{15\xspace}
\newcommand{\rescuePersPages}{10\xspace}
\newcommand{\rescueSeconds}{300\xspace}

\newcommand{\ruleMinutes}{32\xspace}

\newcommand{\ruleSpeedup}{20\xspace}

\usepackage{colortbl}
\usepackage{tikz}
\usepackage{fontawesome5}
\definecolor{cplain}{HTML}{3B7DD8}\definecolor{csym}{HTML}{F29F2E}\definecolor{ccode}{HTML}{3FA66B}\definecolor{csem}{HTML}{8E5BC4}
\definecolor{cimg}{HTML}{E07B91}\definecolor{cnone}{HTML}{B8BEC7}\definecolor{cink}{HTML}{1F2937}\definecolor{cmute}{HTML}{6B7280}
\definecolor{cform}{HTML}{0F766E}\definecolor{cloc}{HTML}{B45309}\definecolor{cred}{HTML}{D6504C}

\newcommand{\pbar}[3]{\tikz[baseline=-0.6ex]{\fill[#3!12] (0,-0.07) rectangle (#1,0.17);\fill[#3] (0,-0.07) rectangle ({#1*#2},0.17);\draw[#3!70,line width=0.35pt] (0,-0.07) rectangle (#1,0.17);}}

\newcommand{\ci}[7]{\tikz[baseline=-0.6ex]{\pgfmathsetmacro\s{#1/(#6-#5)}%
 \draw[black!15,line width=0.3pt] (0,0.05)--(#1,0.05);\draw[black!25] (0,-0.03)--(0,0.13) (#1,-0.03)--(#1,0.13);%
 \draw[#7,line width=0.9pt] ({(#3-#5)*\s},0.05)--({(#4-#5)*\s},0.05);\draw[#7,line width=0.6pt] ({(#3-#5)*\s},-0.02)--({(#3-#5)*\s},0.12) ({(#4-#5)*\s},-0.02)--({(#4-#5)*\s},0.12);%
 \fill[#7] ({(#2-#5)*\s},0.05) circle (1.5pt);}}
\definecolor{cftext}{HTML}{5B8DB8}\definecolor{cfimg}{HTML}{E3A64B}\definecolor{cfact}{HTML}{C25B5B}
\definecolor{clhome}{HTML}{4E79A7}\definecolor{clcontact}{HTML}{59A14F}\definecolor{clcv}{HTML}{9CBF6B}
\definecolor{clsame}{HTML}{8FA3BA}\definecolor{clext}{HTML}{F28E2B}\definecolor{clpdf}{HTML}{B07AA1}
\newcommand{\grouprow}[2]{\rowcolor{black!7}\multicolumn{#1}{@{}l}{\textit{#2}} \\}

\setcopyright{none}
\renewcommand\footnotetextcopyrightpermission[1]{}
\begin{document}

\title[{Beyond ``[at]'': A Large-Scale Measurement of Email Address Presentation}]{Beyond ``[at]'': A Large-Scale Measurement of Email Address Presentation on Computer Science Scholars' Homepages}
\author{Aichun Huang}
\affiliation{\institution{Nankai University}\city{Tianjin}\country{China}}
\author{Xiang Li}
\authornote{Corresponding author: Xiang Li (lixiang(at)nankai.edu.cn).}
\affiliation{\institution{Nankai University}\city{Tianjin}\country{China}}
\author{Zheli Liu}
\affiliation{\institution{Nankai University}\city{Tianjin}\country{China}}
\renewcommand{\shortauthors}{Huang, Li, and Liu}

\begin{abstract}
Publishing an email address makes a scholar reachable, but also exposes it to automated collection.
Yet how email addresses are presented on the Web has not been measured systematically across their content, form, and location.
We measure it across \nTotal homepage records of computer-science scholars listed in CSRankings with \sys, which preserves raw and rendered HTML, extracts addresses with a three-stage cascade (rules, then large language models, then a crawling agent), and labels each address.
Among \nAcc reachable homepages, we find an address for \nEmail scholars (\pEmailAcc).
Our measurement yields four main findings.
\emph{(i)} Presentation is diverse: only \pCPlain publish plain content, the others spread over 23 subtypes of symbol substitution, code, and semantic description, and \pCodeRender of scholars' addresses appear only after rendering.
\emph{(ii)} Presentation follows hosts rather than individuals: scholars on one host share a content class in \pairSameHost of pairs, compared with \pairSameInstDiffHost across hosts of one institution, whereas research output and institution rank are barely related to presentation.
\emph{(iii)} Content and form are coupled through code-based links, and \pLeak of scholars with non-plain selected presentations have plain-text copies of the same address elsewhere.
\emph{(iv)} Plain addresses are more common behind passed bot challenges, and \pLandNonPersonal of redirecting homepages no longer lead to scholars.
Human assessment yields \hvEmailStrict weighted exact-set agreement on \hvNJudged judgeable pages, complemented by three 500-page review rounds with human verification.
Presentation is diverse but clusters by shared host, and by our label-based estimate plain-string matching misses nearly half of the addresses.

\end{abstract}

\maketitle

\section{Introduction}
\label{sec:intro}

Scholars publish email addresses on their homepages to be reachable by collaborators, students, and other readers.
Public contact information is also available to automated collectors, which supply addresses for spam and phishing~\cite{prince2005spammers,shue2009spamology,polakis2010harvest,jagatic2007social}.
To stay reachable while making collection harder, a page may replace the at sign with a word, encode the address in JavaScript, or display it as an image.
Consider one email address published in three ways: as ``\texttt{jdoe@example.edu}'', as ``\texttt{jdoe [at] example [dot] edu}'', and as an ``Email'' icon whose link a script assembles.
A regular expression over the served HTML finds only the first, a lexicon adds the second, and only a collector that executes the page finds the third.
Much of this practice dates from an era of regular-expression harvesters, whereas modern Web agents render pages, follow links, and read text and images~\cite{zhou2024webarena,he2024webvoyager,zheng2024seeact}; and many homepages are now generated by shared templates, so a presentation may reflect infrastructure rather than a personal choice.
We ask how email addresses are presented on scholars' homepages and linked documents, and how page design and hosting infrastructure relate to automated collection.
Answering it is a Web mining problem over multimedia and multilingual content: addresses appear in page source, rendered HTML, images, and linked documents, sometimes described in non-English text.

\noindent\textbf{Prior work and research gap.}
Prior work shows that a published address can be harvested and abused.
Honeypot studies traced collectors with trap addresses~\cite{prince2005spammers} and related incoming spam to where an address was published~\cite{shue2009spamology}, and social networks were shown to expose addresses at scale~\cite{polakis2010harvest}.
On the defense side, controlled experiments compared obfuscation techniques by the spam they attracted~\cite{muehlemann2008nine}, and a Web crawl counted how often they occur~\cite{poornachandran2016obfuscation}; academic search systems, in turn, extract contact details from researcher homepages~\cite{tang2008arnetminer}.
These studies examine either the harvester, a single defense, or the extractor.
A joint characterization is still missing: address content, form, location, cross-page variation, and their links to hosting, access controls, and modern collectors.

\noindent\textbf{Our study.}
We study this question on CSRankings~\cite{berger2026csrankings}, which lists \nTotal homepage records of computer-science scholars at \nInstAll affiliations.
The list defines a community with public homepage URLs and affiliations, so we can compare scholars within and across institutions; we do not assume that every scholar manages the website or chose its presentation.
To measure these pages at scale, we develop \sys, a framework with four modules (Section~\ref{sec:framework}).
It stores each homepage both as the raw HTML returned by the server and as the HTML rendered by a browser, separates reachable homepages from inaccessible and duplicated ones, and extracts addresses through a cascade of rules, LLMs, and a crawling agent that invokes a later layer only when the earlier one finds no address.
Finally, one set of rules assigns content, form, and location labels from the presentation evidence, regardless of which layer found the address.
Two researchers who did not build the pipeline annotated a stratified \hvN-page sample, and a human verified every case in three 500-page rounds of LLM-assisted review (\S\ref{sec:validation}).

\noindent\textbf{Research questions.}
We structure the measurement around four questions, from address presentation to what a reader obtains:
\begin{itemize}
\item \emph{RQ1 (Landscape):} How are scholars' email addresses presented by content, form, and location, and which subtypes occur? (\S\ref{sec:rq1})
\item \emph{RQ2 (Interplay):} How do the dimensions combine, and how does the same address vary in presentation across pages? (\S\ref{sec:rq2})
\item \emph{RQ3 (Ecosystem):} Which shared factors, such as the host, site framework, institution, region, and research output, are associated with the email address presentation? (\S\ref{sec:rq3})
\item \emph{RQ4 (Access):} How do bot management and the decay of registered homepages affect what a reader obtains? (\S\ref{sec:rq4})
\end{itemize}

\noindent\textbf{Our findings.}
Of the \nUnique unique scholars, \nAcc have a reachable homepage, and \nEmail of them (\pEmailAcc) yield an address.

\emph{RQ1: Numerous scholars do not publish a plain address.}
Plain is the most common content class (\pCPlain), but nearly half of the scholars use symbol substitution (\pCSymbolic), code (\pCCode), or a semantic description (\pCSemantic).
Dominant subtypes include \emph{at}/\emph{dot} words (\cSubst), link-only and render-only addresses (\cCode), and name templates (\cSem); rare cases include shell commands and formulas.
Under our capability model, plain-string matching covers \pLadRegex of scholars, rising to \pLadLex with symbol normalization and \pLadJs with rendering and decoding; \pLHomepage of scholars publish on the registered homepage itself.

\emph{RQ2: Content and form are coupled; plain-text copies coexist.}
Content and form are strongly associated (Cram\'er's $V$ = \VCF) because link-only and click-to-reveal addresses are both \cCode and \fAct, whereas location is weakly associated with both (\VCL and \VFL).
Of 72 possible combinations, \combCXFXLObs occur and \combCXFXLEighty cover 80\% of scholars.
Among scholars with non-plain selected presentations, \pLeak have plain-text copies of the same address elsewhere.

\emph{RQ3: Presentation is associated with infrastructure, but barely with prominence.}
Two scholars on the same host share a content class in \pairSameHost of pairs, against \pairSameInstDiffHost across hosts of one institution and \pairExp for random pairs, and \pHostUniform of hosts with four or more scholars are entirely plain or entirely non-plain.
Research portals publish \swResearchPortalNP non-plain addresses and institution-run content management systems only \swUniversityCmsNP; four institution clusters emerge, two of them dominated by \cCode.
Regional differences are substantial (from \npCn in China to \npJp in Japan), while research output and institution rank are barely associated with presentation.

\emph{RQ4: Access controls and link decay change what a reader obtains.}
For \pDoorChal of the homepages, a plain request receives a challenge that our browser passes, and \pDoorLocked remain inaccessible.
Among address-bearing pages reached after a challenge, plain addresses are more common (\pVaultOpenGivenChal) than on pages without one (\pVaultOpenGivenOpen), and a few vendors account for most barriers.
Registered links also decay: \pRedirect of reachable homepages redirect, and \pLandNonPersonal of these no longer lead to a page about the scholar.

\noindent\textbf{Contributions.}
We made the following contributions.
\begin{itemize}
\item \emph{To our knowledge, the first systematic, multi-dimensional measurement} of how email addresses are presented, covering raw and rendered HTML, images, and linked documents for \nEmail scholars, with a content--form--location taxonomy of 23 named subtypes and human-validated labels (\S\ref{sec:background}, \S\ref{sec:rq1}, \S\ref{sec:rq2}).
\item \emph{\sys}, a measurement framework that keeps raw and rendered HTML and extracts email addresses with a cascade of rules, LLMs, and a crawling agent, validated by independent annotators and three review rounds (\S\ref{sec:framework}, \S\ref{sec:data}).
\item \emph{An ecosystem analysis} showing that presentation clusters by host rather than by institution or research output, and relating it to frameworks, regions, bot management, and redirects (\S\ref{sec:rq3}, \S\ref{sec:rq4}).
\end{itemize}

\section{Background and Taxonomy}
\label{sec:background}

\subsection{Background}

\noindent\textbf{Email addresses on the scholarly Web.}
A scholar's homepage presents publications, teaching, contact details, and an academic identity~\cite{hyland2012homepages}; directories such as CSRankings~\cite{berger2026csrankings} link to these pages.
Pages are hand-written, generated, hosted on personal platforms, or produced by institutional content management systems whose templates fix many page elements.

\noindent\textbf{Automated collection.}
Collectors differ in capability.
The simplest fetches the raw HTML and applies a regular expression~\cite{prince2005spammers,polakis2010harvest}; a lexicon can normalize spellings such as ``\texttt{[at]}'', and an HTML parser reads \texttt{mailto} links and entities.
Browser-based crawlers execute JavaScript~\cite{englehardt2016tracking}, optical character recognition reads images~\cite{smith2007tesseract}, and LLM-based Web agents combine link traversal with extraction from text and images~\cite{zhou2024webarena,he2024webvoyager,zheng2024seeact}.
Collected addresses feed spam and phishing~\cite{kanich2008spamalytics,jagatic2007social}, which LLMs now personalize cheaply~\cite{hazell2023spear,heiding2024phishing}.

\noindent\textbf{Protection mechanisms.}
At the address level, a page can spell the address with substitutions, write it as HTML entities, assemble it with a script~\cite{muehlemann2008nine,poornachandran2016obfuscation}, draw it as an image, or put it behind a CAPTCHA~\cite{vonahn2003captcha}, as reCAPTCHA Mailhide did~\cite{vonahn2008recaptcha}.
Content delivery networks offer this for a whole site: Cloudflare's Email Address Obfuscation replaces every address with an encoded string that a script decodes in the browser~\cite{cloudflare2026email}.
At the site level, \emph{bot management} detects automated requests and allows, challenges, or blocks them, for example with browser fingerprints, JavaScript challenges, or access denials~\cite{azad2020webrunner,li2021goodbot,vastel2020fpcrawlers}; it governs access to a page rather than the representation of an address within it.
Separately, \texttt{robots.txt} declares crawling policies for compliant crawlers~\cite{longpre2024consent}.

\noindent\textbf{Which addresses count.}
An address counts if it is the scholar's own or if the page names an assistant, lab, or group address as the scholar's contact; school-wide mailboxes, contact forms, and postal addresses do not count.
A scholar may publish several such addresses, each in several ways; Section~\ref{sec:priority} selects one presentation.

\subsection{Three Dimensions}
\label{sec:dims}

We describe each presentation along three dimensions (Table~\ref{tab:tax}) and assign the labels from the presentation itself, independently of the extraction layer that recovered the address.

\noindent\textbf{Content} describes how the address is represented; the four classes are ordered by the transformation a reader or script must apply.
\emph{(i)} \cPlain displays the full address without substitution or reconstruction; for text, the same string must occur in the served HTML or PDF text, and for images, content follows the displayed characters.
\emph{(ii)} \cSubst replaces some characters by symbols or words that a fixed lexicon maps back: \emph{at}/\emph{dot} words, brackets, separators, look-alike characters, or an icon-font at sign.
\emph{(iii)} \cCode makes the address available through markup, source inspection, decoding, or browser processing: a \texttt{mailto} link or icon, HTML entities, script assembly or decoding, Cloudflare encoding, source-only fields, or text that appears only after rendering.
\emph{(iv)} \cSem specifies the address through a description, a name-based template, an edit instruction, a mask, substitutions outside the lexicon, or an expression displayed as text, which the reader must apply.
The boundary between \cCode and \cSem is who performs the transformation: the browser executing a script (\cCode) or a reader evaluating a displayed command (\cSem).
Labels describe the presentation, not the extractor: a rule may reconstruct a familiar \cSem template, and an LLM may extract a \cPlain string.

\noindent\textbf{Form} describes what a visitor must do.
\emph{(i)} \fText is read directly, including a PDF text layer; \emph{(ii)} \fImage shows the address, or a part such as the at sign, as a picture; \emph{(iii)} \fAct requires an action: clicking a \texttt{mailto} icon or link whose visible text is not the address, clicking to reveal it, solving a CAPTCHA, or opening the page source.
The dimensions combine: an image of an ordinary address is \cPlain and \fImage, and a \texttt{mailto} icon is \cCode and \fAct.

\noindent\textbf{Location} identifies the page or document relative to the registered homepage: \emph{(i)} the homepage, including its redirect destination and automatically loaded frames; \emph{(ii)} a contact page and \emph{(iii)} a CV page on the same site; \emph{(iv)} another page on the same registrable domain~\cite{psl2026}, so \texttt{dbs.example.edu} and \texttt{example.edu} are on the same site; \emph{(v)} a page on another registrable domain; and \emph{(vi)} a linked PDF.
The class orders serve only to select one presentation per scholar (\S\ref{sec:priority}); they are not measured rankings of effort or extraction time.

\begin{table}[t]
\centering
\caption{Classes of the three dimensions in precedence order (Algorithm~\ref{alg:priority}), with schematic examples and the share of the \nEmail address-bearing scholars whose selected presentation has each class. Section~\ref{sec:dims} gives the definitions; no real email address is publicly disclosed.}
\label{tab:tax}
\small
\setlength{\tabcolsep}{3pt}
\begin{tabularx}{\columnwidth}{@{}l>{\raggedright\arraybackslash}Xl@{}}
\toprule
\textbf{Class} & \textbf{Schematic example} & \textbf{Share} \\
\midrule
\grouprow{3}{Content}
\cPlain & \texttt{jane.doe@cs.example.edu} & \pbar{0.55}{0.5378}{cplain}\,{\scriptsize \pCPlain} \\
\cSubst & \texttt{jane.doe [at] cs [dot] example.edu} & \pbar{0.55}{0.2353}{csym}\,{\scriptsize \pCSymbolic} \\
\cCode & an ``Email'' icon linking to \texttt{mailto:jane.doe@...}; \texttt{'jane.doe'+'@'+...} in a script & \pbar{0.55}{0.1855}{ccode}\,{\scriptsize \pCCode} \\
\cSem & ``\texttt{firstname.lastname@\allowbreak example.edu}''; ``my surname, then the at sign, then \texttt{cs.example.edu}'' & \pbar{0.55}{0.0414}{csem}\,{\scriptsize \pCSemantic} \\
\midrule
\grouprow{3}{Form}
\fText & any content example shown as text & \pbar{0.55}{0.8817}{cftext}\,{\scriptsize \pFText} \\
\fImage & a PNG that shows the address & \pbar{0.55}{0.0207}{cfimg}\,{\scriptsize \pFImage} \\
\fAct & ``Email me'' link; ``show email'' button & \pbar{0.55}{0.0976}{cfact}\,{\scriptsize \pFAction} \\
\midrule
\grouprow{3}{Location}
Homepage & \texttt{example.edu/\textasciitilde jdoe/} & \pbar{0.55}{0.8318}{clhome}\,{\scriptsize \pLHomepage} \\
Contact & \texttt{.../\textasciitilde jdoe/contact.html} & \pbar{0.55}{0.0369}{clcontact}\,{\scriptsize \pLContactpage} \\
CV & \texttt{.../\textasciitilde jdoe/cv.html} & \pbar{0.55}{0.0019}{clcv}\,{\scriptsize \pLCVpage} \\
Same-site & \texttt{dbs.example.edu/people/jdoe} & \pbar{0.55}{0.0771}{clsame}\,{\scriptsize \pLOthersamesitepage} \\
External & \texttt{jdoe.github.io} & \pbar{0.55}{0.0165}{clext}\,{\scriptsize \pLOthersite} \\
PDF & \texttt{.../\textasciitilde jdoe/cv.pdf} & \pbar{0.55}{0.0358}{clpdf}\,{\scriptsize \pLPDF} \\
\bottomrule
\end{tabularx}
\end{table}

\subsection{Addresses with Several Presentations}
\label{sec:priority}

A \emph{presentation} is one appearance of an address on one page, possibly combining several techniques; a scholar may publish several addresses and several presentations of each.
Algorithm~\ref{alg:priority} (Appendix~\ref{app:details}) resolves three cases in order.
\emph{(1)} Within one presentation, we assign the highest-ranked content and form labels among its techniques: an address requiring a semantic instruction and a symbol substitution is \cSem, and an image shown only after a click is \fAct.
\emph{(2)} Among the presentations of one address, we select the first available location, then the lowest-ranked content and form labels.
\emph{(3)} Among a scholar's addresses, we select one by comparing location, content, and form in that order, so each scholar contributes one label triple.
When a plain request returns a bot challenge or a redirect stub, we classify the page that a browser obtains and record the barrier separately (\S\ref{sec:rq4}).

\section{The MailScope Framework}
\label{sec:framework}

\begin{figure*}[t]
\centering
\includegraphics[width=\textwidth]{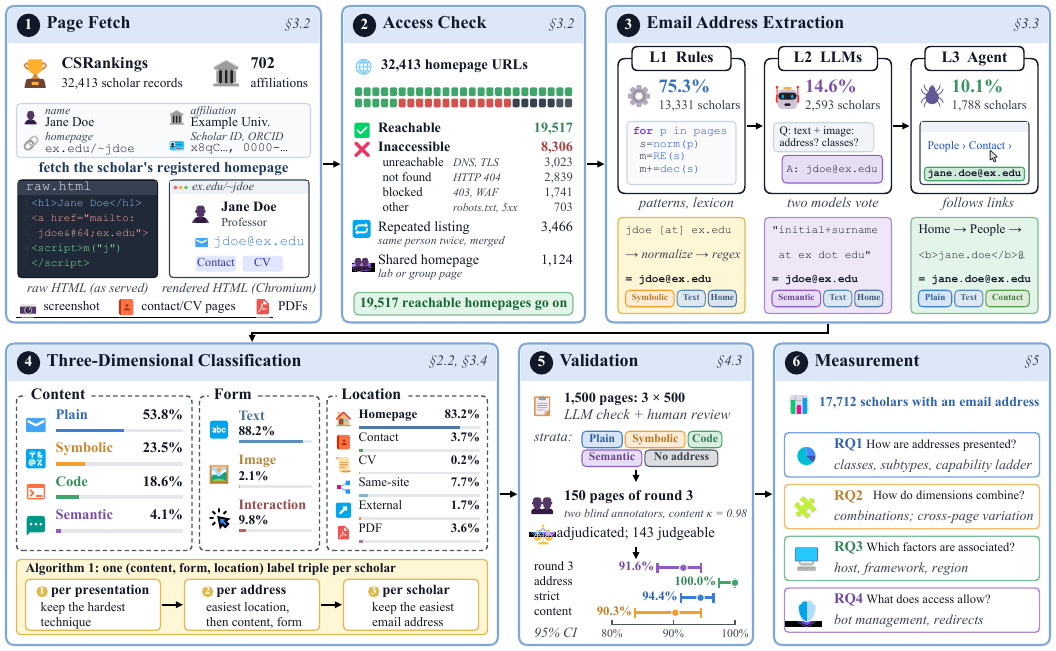}
\Description{Six-panel overview of MailScope: page fetch, access check, three-layer email extraction, three-dimensional classification, validation, and the four research questions.}
\caption{Overview of \sys. Panels cover acquisition, access checks, extraction, classification, validation, and measurement; extraction percentages are shares of the \nEmail address-bearing scholars recovered by each layer.}
\label{fig:arch}
\end{figure*}

\sys turns a list of registered homepages into a classified set of email addresses through four modules (Figure~\ref{fig:arch}): page fetch (\S\ref{sec:acquire}), accessibility check (\S\ref{sec:acquire}), email address extraction (\S\ref{sec:extract}), and three-dimensional classification (\S\ref{sec:classify}).
We assess the output through human annotation and LLM-assisted review (\S\ref{sec:valproto}).

\subsection{Design Overview}
\label{sec:design}

We keep both the served and the rendered HTML, because they can contain different presentations (a script-built address appears only after rendering), and we must handle images and natural-language descriptions without calling a model on every page.
We therefore apply the deterministic rules first, LLMs to the unresolved pages, and a crawling agent only when no email address has been recovered~\cite{chen2023frugalgpt}; the rule layer runs about \ruleSpeedup times faster than the LLM layer (Appendix~\ref{app:extract}).

\subsection{Page Fetch}
\label{sec:acquire}

CSRankings lists computer-science scholars with an affiliation and a homepage URL and identifies them by their DBLP names~\cite{berger2026csrankings,ley2009dblp}; its author lists give \nTotal records.
For each homepage, we store the raw HTML that the server returns, the HTML rendered by headless Chromium after the page's scripts have run, and a screenshot.
We also fetch linked contact pages, CV pages, and personal websites, and extract text from linked CV PDFs; when a homepage redirects to another site, we fetch up to three contact or CV links on the destination site.
Keeping both HTML versions separates an address sent by the server from one that appears only after scripts run, in the spirit of rendering-based Web measurement~\cite{englehardt2016tracking,jueckstock2021realistic}.
We space requests to one host by a random delay of 1--2.5 seconds; the homepages span \nHosts hosts, with a median of one page per host.
We exclude homepages disallowed by \texttt{robots.txt} and treat the September 2026 crawl as one snapshot.
\label{sec:check}An accessibility check then labels each record as reachable, repeated or shared, or one of seven failure types (e.g., unreachable host, not found, blocked); only reachable homepages are analyzed (Section~\ref{sec:data}).

\subsection{Email Address Extraction}
\label{sec:extract}

The rule layer (L1) processes every reachable page, the LLM layer (L2) the pages on which L1 finds no confident own address, and the crawling agent (L3) the pages on which L2 also finds nothing.

\noindent\textbf{L1: Rules.}
Regular expressions, a multilingual substitution lexicon, and parsers for \texttt{mailto} links, HTML entities, Cloudflare encoding, and JavaScript assembly patterns cover \cPlain, \cSubst, and most \cCode presentations; an address present in the rendered but not the raw HTML is marked as render-only.
Ownership signals (name matches, role terms, generic mailbox names, and addresses repeated across scholars' pages) assign each candidate to its owner.

\noindent\textbf{L2: LLMs.}
Two LLMs from different providers, Claude Sonnet 5.5 and GLM-5.3, read the page text, the HTML around email clues, linked pages, and vision-model transcriptions of images; we accept matching outputs, and Claude Opus 5.5 resolves disagreements.

\noindent\textbf{L3: Crawling agent.}
A Claude Sonnet 5.5 agent follows links and runs site searches for the scholar's name within page and time budgets, without general Web search, CAPTCHA solving, authentication, or bypassing bot checks.
Appendix~\ref{app:extract} details the rules, inputs, budgets, and limits of the three layers.

\subsection{Three-Dimensional Classification}
\label{sec:classify}

We classify each address from the archived evidence (raw and rendered HTML, linked pages, documents retrieved by L2 and L3, and PDF text) with one set of rules, independently of the extraction layer.
For each address, we record every presentation with its content, form, and location (\S\ref{sec:dims}) and apply Algorithm~\ref{alg:priority}.
A rule-based reconstruction of a template keeps its \cSem label, and using an LLM for extraction does not by itself imply \cSem content.
For \pFallback of the selected records, the archive lacks enough evidence to derive the label, for example when a template was processed on a fetched page that was not preserved; we then retain the label assigned during extraction.
This keeps the taxonomy stable when a layer changes; Appendix~\ref{app:details} details the rules.

\subsection{Validation Design}
\label{sec:valproto}

We draw three non-overlapping stratified samples of 500 pages; Opus gives an initial check of each address set and its labels, and a human reviewer verifies every case and makes the final call.
From round~3, two researchers who did not build the pipeline blindly annotate a \hvN-page subset (30 pages per content class and 30 without an address), an author adjudicates disagreements, and we report weighted agreement, 95\% intervals, and Cohen's $\kappa$~\cite{cohen1960kappa} (Appendix~\ref{app:estimation}).
We report the rounds separately because each evaluates the pipeline version in use at that time.
Strata are the four content classes and the no-address group with equal allocation, so rare classes such as \cSem receive as many reviewed pages as \cPlain, and weights restore population shares (Appendix~\ref{app:estimation}).

\section{Dataset and Pipeline Evaluation}
\label{sec:data}

\subsection{Dataset}
\label{sec:dataset}

From the \nTotal CSRankings records, we retain \nEmail address-bearing scholars for analysis (Table~\ref{tab:funnel}).

\begin{figure*}[t]
\centering
\includegraphics[width=\textwidth]{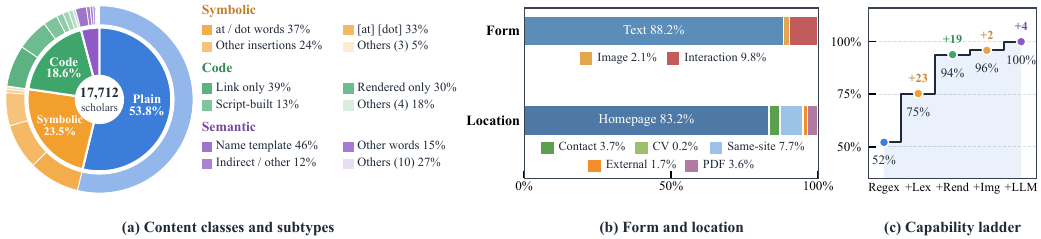}
\Description{Donut chart of content classes and subtypes, stacked bars of form and location shares, and a step chart of estimated coverage as extraction capabilities are added.}
\caption{Distribution of the \nEmail address-bearing scholars by their selected address presentation. (a) Content classes (inner ring) and their subtypes (outer ring); (b) form and location; (c) estimated share of scholars covered by successive extraction capabilities, based on presentation labels (document access and available PDF text are assumed).}
\label{fig:dist}
\end{figure*}

\begin{table}[t]
\centering
\caption{From CSRankings records to analyzed email addresses. Shares use the unique scholars as the base, except for the last two rows, which use the reachable homepages.}
\label{tab:funnel}
\small
\begin{tabularx}{\columnwidth}{@{}Xrl@{}}
\toprule
\textbf{Stage} & \textbf{Records} & \textbf{Share} \\
\midrule
\cellcolor{black!7}CSRankings records & \cellcolor{black!7}\nTotal & -- \\
\quad repeated listings of one scholar (merged) & \nDupSame & -- \\
\midrule
\cellcolor{black!7}Unique scholars & \cellcolor{black!7}\nUnique & \pbar{0.55}{1.0}{cplain}\,{\scriptsize 100\%} \\
\quad homepage shared with another scholar & \nSharedUnique & \pbar{0.55}{0.039}{csym}\,{\scriptsize \pSharedUnique} \\
\quad homepage inaccessible & \nInacc & \pbar{0.55}{0.287}{cred}\,{\scriptsize \pInacc} \\
\midrule
\cellcolor{black!7}Reachable homepages & \cellcolor{black!7}\nAcc & \pbar{0.55}{0.674}{cplain}\,{\scriptsize \pAcc} \\
\quad with an email address (analyzed) & \nEmail & \pbar{0.55}{0.908}{cplain}\,{\scriptsize \pEmailAcc} \\
\quad no email address found & \nNoEmailAcc & \pbar{0.55}{0.092}{cred}\,{\scriptsize \pNoEmailAcc} \\
\bottomrule
\end{tabularx}
\end{table}

Merging \nDupSame repeated listings leaves \nUnique unique scholars; \nSharedUnique share a URL, usually a lab page.
We exclude \nInacc inaccessible homepages (\pInacc), mostly because the host does not respond (\nSUnreachable), the page is gone (\nSNotFound), or our crawler is blocked (\nSBlocked).
Of the \nAcc reachable homepages, \nEmail (\pEmailAcc) yield an address; the remaining \nNoEmailAcc are an upper bound on scholars without a published address (Appendix~\ref{app:data}).
The records cover \nInstAll affiliations; institution-level summaries use the \nInstBig institutions with at least five address-bearing scholars, and the clustering in \S\ref{sec:rq3} uses those with at least 15.

\subsection{Contribution of the Extraction Layers}
\label{sec:layers}

L1 recovers the addresses of \pSrcRule of address-bearing scholars, L2 of \pSrcLlm, and L3 of \pSrcRescue (Table~\ref{tab:stage} in the appendix); L2 records contain more \cSem addresses (17.3\%) than L1 records (0.7\%), consistent with routing unresolved presentations onward.
L3 searched the \nRescueIn reachable homepages on which L1 and L2 had found nothing and recovered \nSrcRescue addresses.

\subsection{Validation Results}
\label{sec:validation}

\noindent\textbf{LLM-assisted human review.}
The three 500-page rounds give weighted address-set agreement of 92.2\% [88.4, 94.8], 91.1\% [87.0, 94.0], and 91.6\% [87.4, 94.5]; round~3 evaluates the frozen pipeline.
Correct no-address decisions count as agreement, and missed, partial, and wrong extractions count as errors.
On cases with a correct address, the label agreement across the rounds is 94.9--96.2\% for content, 96.8--98.1\% for form, and 96.1--98.2\% for location.

\noindent\textbf{Independent double annotation (\hvN sampled; \hvNJudged evaluable).}
The evaluated groups contain 30 pages for each content class and 23 no-address pages.
The two annotators agree on address presence for \hvAgreeHas of pages ($\kappa=\hvKappaHas$) and on content for \hvAgreeContent ($\kappa=\hvKappaContent$).
After adjudication, complete address sets match on \hvStrictGoodN of \hvNJudged pages (unweighted \hvEmailStrictRaw; weighted \hvEmailStrict, 95\% CI \hvEmailStrictCI); the remaining cases consist of \hvMissN addresses found only by the annotators and \hvExtraN found only by the pipeline.
On the \hvContentN evaluable address-bearing cases, content, form, and location agree in \hvContent, \hvForm, and \hvLocation; content agreement ranges from 86.2\% for \cSem and 86.7\% for \cSubst to 96.7\% for \cPlain.

\section{Results}
\label{sec:results}

We analyze the \nEmail scholars with an address from a reachable homepage or a linked document; unless stated otherwise, each scholar contributes the presentation selected by Algorithm~\ref{alg:priority}, and subgroup percentages use the stated subgroup as denominator.

\subsection{RQ1: How Are Email Addresses Presented?}
\label{sec:rq1}

\noindent\textbf{Plain content and homepage publication are most common.}
\cPlain accounts for \pCPlain of the scholars, \cSubst for \pCSymbolic, \cCode for \pCCode, and \cSem for \pCSemantic (Figure~\ref{fig:dist}a).
For form (Figure~\ref{fig:dist}b), \pFText are \fText, \pFAction \fAct, and \pFImage \fImage.
For location, \pLHomepage have the selected presentation on the registered homepage, including its redirect destination and automatically loaded frames.
For the remaining \pAwayHome (\nAwayHome scholars), we recover an address only from another page or document, such as a contact page (\pLContactpage), another same-site page (\pLOthersamesitepage), or a PDF (\pLPDF).

\noindent\textbf{Rendering and decoding increase estimated coverage.}
We map the content and form labels to the capabilities needed to recover each address (Figure~\ref{fig:dist}c).
This is a label-based estimate, not a run of separate extractors, and it assumes that the relevant document has been retrieved.
Plain-string matching covers \pLadRegex of scholars (1.7\% show \cPlain addresses only as images), symbol normalization \pLadLex, rendering and decoding \pLadJs, and image-to-text conversion \pLadVision.
The remaining \pNeedLlm require reconstruction from descriptions or instructions.
The estimate does not measure navigation success or the ability to bypass access controls.

\begin{figure*}[t]
\centering
\includegraphics[width=\textwidth]{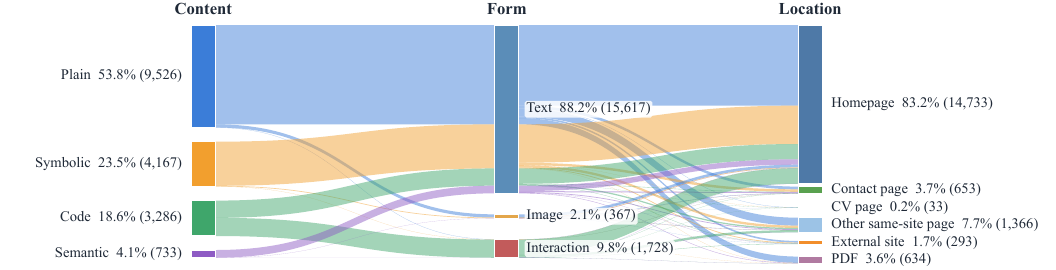}
\Description{Sankey diagram linking each scholar's content class to form and location.}
\caption{Distribution of the \nEmail address-bearing scholars across content (left), form (middle), and location (right). Each scholar contributes one selected label triple; ribbon width represents the number of scholars.}
\label{fig:flow}
\end{figure*}

\noindent\textbf{Each non-plain class has a few dominant subtypes.}
Among the \nCCode \cCode scholars, \pSubMailto have an address available only through a \texttt{mailto} link or icon, and \pSubRender only after rendering (\pCodeRender of all scholars, invisible to a source-only crawl); these ordinary page mechanisms do not by themselves indicate an intention to hide the address.
Scripts assemble or encode \pSubEncoded, HTML entities account for \pSubEntity, source-only strings (comments, attributes, hidden elements) for \pSubSourceOnly, and Cloudflare's encoding for \pSubCloudflare.
Within \cSubst, \psymWords use the words \emph{at} and \emph{dot}, \psymBr put them in brackets, and \psymOther insert separators that leave the address readable; look-alike characters are rare (\psymChar).
Within \cSem, the name template dominates (\PSubSemTemplate), as in ``\texttt{firstname.lastname}'' or ``my last name'', followed by substitutions outside the lexicon (\PSubSemWordSubstitution), descriptions (\PSubSemDescription), and edit instructions (\PSubSemInstruction).
Of the word-based \cSem addresses, \pSemNonEn are not in English, for example a French \emph{chez} or a Dutch \emph{apestaartje} for the at sign.

\noindent\textbf{Rare presentations require additional transformations.}
The non-plain classes span 23 named subtypes (Figure~\ref{fig:sub}); some scholars write the address as a shell command or a formula, spell it with emoji, or insert hidden words (``\texttt{[CTRL-C is disabled]}'') so that a copied string is wrong; others let a script compute the at sign or a style rule restore a reversed address (Table~\ref{tab:cases} in the appendix).

\begin{takeaway}
Only \pCPlain publish plain content, so plain-string matching covers \pLadRegex of scholars; symbol normalization, rendering, and decoding raise coverage to \pLadJs, and \pLHomepage publish on the registered homepage itself.
\end{takeaway}

\subsection{RQ2: How Do Dimensions Combine and Presentations Vary?}
\label{sec:rq2}

\noindent\textbf{Content and form are coupled through \cCode.}
We measure pairwise associations with Cram\'er's $V$~\cite{cramer1946methods}, which ranges from zero (no association) to one (maximal association).
Content and form give $V=\VCF$ (\VCFCI), whereas content and location give only \VCL and form and location \VFL (Figure~\ref{fig:flow}).
The coupling comes from \cCode: \pCFCodeAction of the \cCode addresses need an interaction, because a \texttt{mailto} icon or a click-to-reveal button is both code and interaction, while more than 96\% of the addresses in every other class are text.

\noindent\textbf{A few combinations dominate.}
Of the 72 possible combinations of content, form, and location, \combCXFXLObs occur, and \combCXFXLEighty cover 80\% of the scholars.
Plain--Text--Homepage alone accounts for \pTopTriple, and the three most common combinations together for \pTopThreeTriples.
\cCode has the highest homepage share (\pCLCodeHomepage), as it relies on homepage markup or scripts, and \cSem the lowest (\pCLSemanticHomepage), with \pCLSemanticContactpage on a contact page where explanations can be given separately.
An extractor restricted to homepages would miss the \pAwayHome of scholars whose addresses appear only elsewhere.

\noindent\textbf{Non-plain presentations coexist with plain-text copies.}
Of the \nNonPlain scholars whose selected presentation is non-plain, \nLeak (\pLeak) show the same address in plain text on another page or in a PDF, so a crawler that reaches that page needs no decoding.
In addition, \pTwoAddr of the scholars publish two own addresses and \pThreePlusAddr three or more, which gives a collector several chances to find one.

\begin{takeaway}
Content and form are coupled through \cCode.
Among scholars with non-plain selected presentations, \pLeak have plain-text copies of the same address elsewhere.
\end{takeaway}

\subsection{RQ3: What Is Associated with Presentation?}
\label{sec:rq3}

\begin{figure*}[t]
\centering
\includegraphics[width=\textwidth]{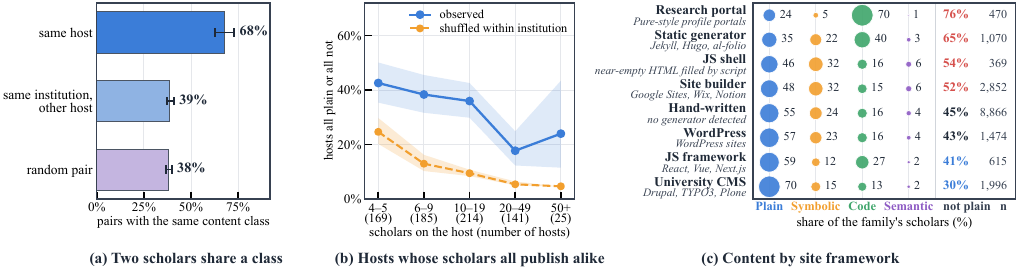}
\Description{Bar chart of same-class pair shares by host relation, line chart of uniform hosts by host size versus a shuffled baseline, and bubble chart of content by site framework.}
\caption{(a) Share of scholar pairs with the same content class (95\% intervals from resampling institutions). (b) Share of hosts whose scholars are all plain or all non-plain, by host size, as observed and with labels shuffled within institutions (bands: 95\% Wilson intervals and the 2.5th--97.5th percentiles of 300 shuffles). (c) Content distribution and non-plain share by homepage framework family (families with at least 200 address-bearing scholars).}
\label{fig:inst}
\end{figure*}

\noindent\textbf{Scholars on one host have similar presentations.}
Two scholars on the same host share a content class in \pairSameHost of pairs (\pairSameHostCI), two scholars of one institution on different hosts in \pairSameInstDiffHost (\pairSameInstDiffHostCI), and two random scholars in \pairExp (Figure~\ref{fig:inst}a; intervals from resampling institutions).
Affiliation alone thus adds little beyond the random baseline, whereas a shared host does.
The intraclass correlation~\cite{shrout1979icc} of ``not \cPlain'' is \iccHost across the \nHostsICC hosts with two or more scholars, against \iccInstDeep across institutions.
Among the \nHostsGEFour hosts with four or more scholars, \pHostUniform contain only \cPlain or only non-plain scholars, compared with \pHostUniformNull when labels are shuffled within institutions.
The observed share exceeds the shuffled share in every host-size group (Figure~\ref{fig:inst}b), so the pattern is not limited to small hosts, on which agreement by chance is more likely.

\begin{figure*}[t]
\centering
\includegraphics[width=\textwidth]{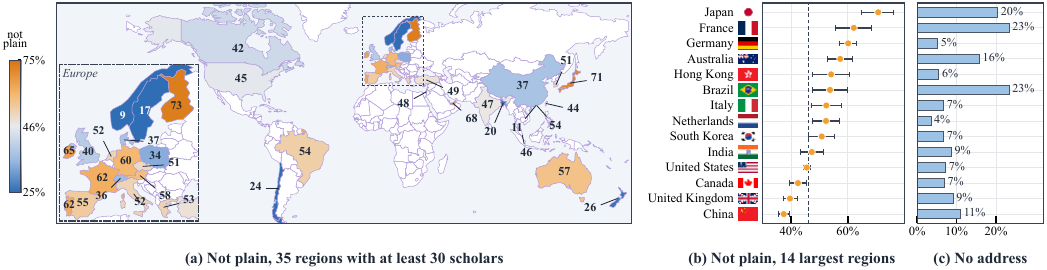}
\Description{World map of non-plain shares by region with a Europe inset, interval plot for the 14 largest regions, and bars of the share without a recovered address.}
\caption{Share of address-bearing scholars with non-plain selected presentations by region of the institution: (a) regions with at least 30 scholars; (b) the 14 largest regions, with 95\% intervals. (c) Share of reachable scholars without a recovered address.}
\label{fig:country}
\end{figure*}

\noindent\textbf{Site frameworks and services concentrate presentations.}
We identify each homepage's framework from fingerprints in its source and group the frameworks into families (Figure~\ref{fig:inst}c).
Research portals, which commonly encode the \texttt{mailto} link, have \swResearchPortalNP non-plain addresses among their \swResearchPortalN scholars, \swResearchPortalCode in \cCode.
Institution-run content management systems have the lowest non-plain share (\swUniversityCmsNP), site builders such as Google Sites a higher one (\swSiteBuilderNP), and WordPress lies near the corpus average (\swWordpressNP).
Site-wide services are also concentrated: Cloudflare's email protection covers \sigNCf scholars at \sigInstCf institutions (\sigNPCf non-plain), and obfuscation plugins cover \sigNPlugin scholars (\sigNPPlugin non-plain).

\noindent\textbf{Four institutional presentation profiles.}
Clustering the \nInstFifteen institutions with at least 15 address-bearing scholars by their combination shares yields four profiles (Appendix~\ref{app:clusters}): a \emph{plain} cluster (\icPlainInst institutions) and a \emph{mixed} cluster (\icMixInst), and two small \cCode-dominated clusters that differ in form, \emph{rendered/encoded} (address text after rendering or decoding) and \emph{link-only} (an ``Email'' link or control).
In these two clusters, the largest host holds on average \icRenConc and \icMailConc of the scholars, consistent with shared profile templates.

\noindent\textbf{Presentation shares vary across regions.}
Regional differences are substantial (Figure~\ref{fig:country}; $\chi^2$ test $p$ \chiCountryP): non-plain addresses are most common in Japan (\npJp), France (\npFr), and Germany (\npDe), and least common in China (\npCn), the United Kingdom (\npGb), and Canada (\npCa).
We recover no address for \naJp of reachable scholars in Japan, compared with \naNl in the Netherlands.
A logistic model of ``not \cPlain'' with research output, ACM Fellow status, and institution rank reaches a McFadden pseudo-$R^2$~\cite{mcfadden1974logit} of only \pRMOne, rising to \pRMFour when region, hosting, and framework are added, compared with \pRInstFE for institution identity alone.

\begin{takeaway}
Presentation varies with hosting, but barely with research output (host intraclass correlation \iccHost vs.\ \iccInstDeep for institutions).
\end{takeaway}

\subsection{RQ4: How Do Site Access and Redirects Affect What a Reader Obtains?}
\label{sec:rq4}

\begin{figure*}[t]
\centering
\includegraphics[width=\textwidth]{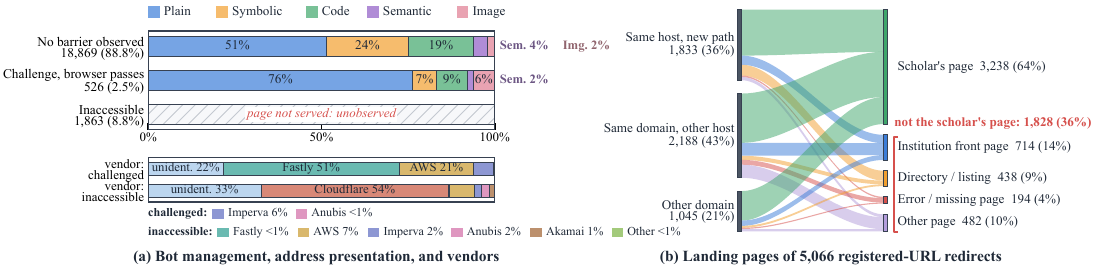}
\Description{Stacked bars of address content by bot-management outcome with vendor breakdowns, and a Sankey diagram of redirect types to landing pages.}
\caption{(a) Top: content of the addresses found behind each kind of site-level bot management (rows give the number and share of homepages; shares may not sum to 100\% because of rounding); address presentation is unobserved on inaccessible pages. Bottom: bot-management products behind challenges that a browser passes and behind inaccessible pages. (b) Landing pages of the \nRedirect redirecting registered URLs, by redirect type; the bracket groups the landing pages that are not the scholar's.}
\label{fig:rq4}
\end{figure*}

\noindent\textbf{Accessible challenged pages more often expose plain addresses.}
Site-level bot management decides whether a client can access a page, whereas address-level obfuscation changes the representation within an accessible page (Figure~\ref{fig:rq4}a, top).
Of the 21,258 homepages that load or block our crawler, \pDoorOpen show no effective barrier.
For \pDoorChal (\nDoorChal pages), a plain HTTP request returns a challenge that browser execution passes; the remaining \pDoorLocked stay inaccessible, \nLockedRefused with an outright refusal across \nBlockedHosts hosts and \nLockedBrowser with an unresolved browser challenge.
Among address-bearing pages reached after a challenge, \pVaultOpenGivenChal of the addresses are plain, compared with \pVaultOpenGivenOpen on pages without an observed barrier.
Because representation on inaccessible pages is unobservable, this comparison does not describe every site that deploys bot management.

\noindent\textbf{A few vendors provide the bot management.}
Most observed challenges come from Fastly (\nChalFastly homepages) and AWS (\nChalAws) (Figure~\ref{fig:rq4}a, bottom).
Among inaccessible homepages, Cloudflare is the most common identified product (\nLockCloudflare), followed by Anubis (\nLockAnubis), AWS (\nLockAws), Imperva (\nLockImperva), and Akamai (\nLockAkamai); we identify the product for \pLockedIdent of inaccessible homepages.
Challenges also cluster by institution: the \nBarrierFixed scholars whose raw HTML is a challenge page belong to \nBarrierInst institutions, and the three most affected institutions account for \pBarrierTopThree of them, so the contrast may partly reflect institutional templates.

\noindent\textbf{More than a third of redirects lead to non-personal pages.}
\pRedirect of reachable registered URLs (\nRedirect) redirect: \nRedSameHost to another path on the same host, \nRedSameDom to another host of the same domain, and \nRedOtherDom to another domain (Figure~\ref{fig:rq4}b).
\pLandNonPersonal of the redirects (\nLandNonPersonal) land on a page that is not the scholar's: an institution's front page (\pLandFront), a directory (\pLandDirectory), an error page (\pLandError), or another page (\pLandOther).
For \pClickLand of address-bearing scholars (\nClickLand), the selected presentation requires following a link from the registered page, most often to a same-host page (\pClickSameHost) or a PDF (\pClickPdf).

\begin{takeaway}
Plain addresses are more common behind passed challenges (\pVaultOpenGivenChal), and \pLandNonPersonal of redirects no longer reach the scholar.
\end{takeaway}

\section{Discussion}
\label{sec:discussion}

\noindent\textbf{Implications for Web measurement.}
For \pCodeRender of address-bearing scholars, the address appears only after rendering, while rendered HTML alone loses server-side encoding evidence, so measurements should keep both views~\cite{englehardt2016tracking,zeber2020representativeness}.
Images, PDFs, and non-English descriptions call for extraction across formats and languages.

\noindent\textbf{Implications for scholars, operators, and directories.}
Presentation is more similar within hosts than across hosts of one institution, suggesting that exposure reviews should include shared templates and site services.
Directories such as CSRankings should periodically check that registered URLs still reach personal pages, because lapsed or dangling domains can be taken over~\cite{lauinger2017registrars,liu2016dangling}.

\noindent\textbf{Implications for security and privacy.}
Our label-based estimates suggest that obfuscation offers limited resistance to collectors that render, decode, and follow instructions~\cite{kim2025online,heiding2024phishing}.
Contact forms or role addresses separate public contact from personal mailboxes.

\noindent\textbf{Limitations.}
Independent double annotation covers \hvNJudged evaluable pages, and the 500-page rounds are LLM-assisted, so the initial assessment may anchor the reviewer, yet blind annotators agree similarly (\hvEmailStrict vs.\ 91--92\%).
Our shares describe reachable scholars with a recovered address, so inaccessible pages are unobserved and ``no address found'' need not mean that none is published; most inaccessible homepages, however, no longer exist and so hide no current presentation.
Hard presentations (images, instructions) are the most likely to be missed, so \cSem and \fImage shares are likely underestimated.
\cSem subtypes come from an LLM classifier, the study covers one community and one snapshot, and host and framework associations are not causal effects.

\noindent\textbf{Ethics.}
We analyze public pages and show only schematic examples.
Our User-Agent named the study and a contact address, and the scanning IP served a website explaining the research purpose with an opt-out; we received no opt-out requests or complaints.
We crawled slowly, honored \texttt{robots.txt}, never bypassed access controls, and release no addresses (Appendix~\ref{app:ethics}).
Lacking an IRB, our institution's equivalent ethics body approved the study; page archives are deleted after a fixed retention period.
We are notifying scholars whose homepages are inaccessible or taken over (e.g., domains re-registered for gambling); several have thanked us.

\section{Related Work}
\label{sec:related}

\noindent\textbf{Harvesting and obfuscation.}
Prior work traces collectors~\cite{prince2005spammers}, relates spam to publication sites~\cite{shue2009spamology}, and measures social-network exposure~\cite{polakis2010harvest}, spam, and phishing~\cite{kanich2008spamalytics,jagatic2007social}.
Experiments and crawls compare obfuscation~\cite{muehlemann2008nine,poornachandran2016obfuscation}; CAPTCHA~\cite{vonahn2008recaptcha} and CDN services~\cite{cloudflare2026email} protect addresses, while automated CAPTCHA solvers improve~\cite{searles2023captcha}.
Unlike counts of obfuscation in served source~\cite{poornachandran2016obfuscation}, we label each address in three dimensions across views, with human validation.

\noindent\textbf{LLMs, crawling, and bot detection.}
LLMs leak personal information from training data~\cite{carlini2021extracting,lukas2023pii}, infer personal attributes~\cite{staab2024memorization}, and collect it from the Web~\cite{kim2025online}.
Crawl configuration affects observations~\cite{englehardt2016tracking,zeber2020representativeness,jueckstock2021realistic}; we relate bot detection~\cite{vastel2020fpcrawlers,azad2020webrunner,li2021goodbot} and crawler restrictions~\cite{longpre2024consent,liu2025somesite,steinacker2026robots} to address presentation.

\noindent\textbf{Web information extraction and decay.}
Prior work extracts structured data through wrappers~\cite{chang2006survey}, profile mining~\cite{tang2008arnetminer}, LLMs~\cite{arora2023evaporate}, and Web agents~\cite{zhou2024webarena,he2024webvoyager}; we measure presentation mechanisms.
Like other Web references~\cite{barYossef2004decay,klein2014rot}, registered homepages decay.

\section{Conclusion}
\label{sec:conclusion}

We measure how \nEmail CSRankings scholars present email addresses with human-validated labels.
Presentation is diverse (only \pCPlain plain, 23 subtypes, \pCodeRender visible only after rendering) and clusters within hosts rather than by institution or research output.
Future work includes real-collector runs and repeated crawls.

\clearpage
\begin{table*}[t]
\centering
\caption{Presentations found in the corpus, in schematic form (names and domains replaced, structure kept).}
\label{tab:cases}
\small
\begin{tabularx}{\textwidth}{@{}p{2.75cm}p{2.1cm}>{\raggedright\arraybackslash}X p{3.9cm}@{}}
\toprule
\textbf{Class} & \textbf{Subtype} & \textbf{Schematic form} & \textbf{What a reader or script must do} \\
\midrule
\multirow{2}{*}{\cSubst} & words & \texttt{jane.doe [at] cs [dot] example [dot] edu} & replace the words by symbols \\
 & emoji & \texttt{[emoji letters]@example.edu} & map the emoji to letters \\
\midrule
\multirow{9}{*}{\cSem} & template & \texttt{firstname.lastname@example.edu}; ``my last name at example.edu'' & fill in the scholar's name \\
 & other words & \texttt{jane.doe[chez]univ-example[point]fr}; \texttt{jane.doe\_ALT\_example.edu} & know what the token stands for \\
 & instruction & \texttt{jane.doe@cs.REMOVE.example.edu}; ``add \texttt{.example.edu}'' & edit the string as told \\
 & decoy words & \texttt{j[CTRL-C is disabled]doe@example.edu} & ignore the inserted words \\
 & alternation & \texttt{jdoe@\{example.com|example.edu\}} & choose a domain \\
 & split pieces & \texttt{| jane.doe | @ | example.org |} (three table cells) & join the pieces \\
 & code expression & \texttt{echo @example.edu | sed 's/\textasciicircum{}/jdoe/'} & evaluate the command \\
 & math & a formula whose value gives a character of the address & compute it \\
 & description & ``first name, a period, family name, at example dot edu'' & build the address \\
\midrule
\multirow{4}{*}{\cCode} & assembled & \texttt{'jdoe' + String.fromCharCode(1{<}{<}6) + 'example.edu'} & run the script \\
 & encoded & \texttt{data-md5="bWFpbHRv..."} and a function that fills the link & decode the attribute \\
 & reversed & \texttt{ude.elpmaxe@eod\_j} restored by a script or a style rule & reverse the string \\
 & rendered only & the raw HTML is an empty shell; the address appears after scripts run & render the page \\
\midrule
Form: \fImage & server-drawn & a server script draws the address as an image for each visit & read the image \\
Form: \fAct & CAPTCHA & a CAPTCHA-gated link that shows only the first letter and the domain & solve the CAPTCHA \\
\bottomrule
\end{tabularx}
\end{table*}

\bibliographystyle{ACM-Reference-Format}
\bibliography{references}

@article{arora2023evaporate,
  title     = {Language Models Enable Simple Systems for Generating Structured Views of Heterogeneous Data Lakes},
  author    = {Arora, Simran and Yang, Brandon and Eyuboglu, Sabri and Narayan, Avanika and Hojel, Andrew and Trummer, Immanuel and R{\'e}, Christopher},
  journal   = {Proceedings of the VLDB Endowment (PVLDB)},
  volume    = {17},
  number    = {2},
  pages     = {92--105},
  year      = {2023},
  doi       = {10.14778/3626292.3626294}
}

@inproceedings{azad2020webrunner,
  title     = {Web Runner 2049: Evaluating Third-Party Anti-bot Services},
  author    = {Amin Azad, Babak and Starov, Oleksii and Laperdrix, Pierre and Nikiforakis, Nick},
  booktitle = {Conference on Detection of Intrusions and Malware, and Vulnerability Assessment (DIMVA)},
  series    = {LNCS},
  volume    = {12223},
  pages     = {135--159},
  year      = {2020},
  doi       = {10.1007/978-3-030-52683-2_7}
}

@inproceedings{barYossef2004decay,
  title     = {Sic Transit Gloria Telae: Towards an Understanding of the Web's Decay},
  author    = {Bar-Yossef, Ziv and Broder, Andrei Z. and Kumar, Ravi and Tomkins, Andrew},
  booktitle = {International Conference on World Wide Web (WWW)},
  pages     = {328--337},
  year      = {2004},
  doi       = {10.1145/988672.988716}
}

@misc{berger2026csrankings,
  title        = {{CSRankings}: CS Rankings},
  author       = {Berger, Emery},
  howpublished = {\url{https://csrankings.org}},
  year         = {2026}
}

@inproceedings{carlini2021extracting,
  title     = {Extracting Training Data from Large Language Models},
  author    = {Nicholas Carlini and Florian Tram{\`e}r and Eric Wallace and Matthew Jagielski and Ariel Herbert-Voss and Katherine Lee and Adam Roberts and Tom Brown and Dawn Song and {\'U}lfar Erlingsson and Alina Oprea and Colin Raffel},
  booktitle = {USENIX Security Symposium (USENIX Security)},
  pages     = {2633--2650},
  year      = {2021}
}

@article{chang2006survey,
  title     = {A Survey of Web Information Extraction Systems},
  author    = {Chang, Chia-Hui and Kayed, Mohammed and Girgis, Moheb Ramzy and Shaalan, Khaled F.},
  journal   = {IEEE Transactions on Knowledge and Data Engineering (TKDE)},
  volume    = {18},
  number    = {10},
  pages     = {1411--1428},
  year      = {2006},
  doi       = {10.1109/TKDE.2006.152}
}

@article{chen2023frugalgpt,
  title   = {{FrugalGPT}: How to Use Large Language Models While Reducing Cost and Improving Performance},
  author  = {Chen, Lingjiao and Zaharia, Matei and Zou, James},
  journal = {arXiv preprint arXiv:2305.05176},
  year    = {2023}
}

@misc{cloudflare2026email,
  title        = {Email Address Obfuscation},
  author       = {{Cloudflare}},
  howpublished = {Cloudflare Web Application Firewall documentation, \url{https://developers.cloudflare.com/waf/tools/scrape-shield/email-address-obfuscation/}},
  year         = {2026},
  note         = {Accessed October 2026}
}

@article{cohen1960kappa,
  title     = {A Coefficient of Agreement for Nominal Scales},
  author    = {Cohen, Jacob},
  journal   = {Educational and Psychological Measurement},
  volume    = {20},
  number    = {1},
  pages     = {37--46},
  year      = {1960},
  doi       = {10.1177/001316446002000104}
}

@book{cramer1946methods,
  title     = {Mathematical Methods of Statistics},
  author    = {Cram{\'e}r, Harald},
  publisher = {Princeton University Press},
  address   = {Princeton, NJ},
  year      = {1946}
}

@inproceedings{englehardt2016tracking,
  title     = {Online Tracking: A 1-million-site Measurement and Analysis},
  author    = {Englehardt, Steven and Narayanan, Arvind},
  booktitle = {ACM SIGSAC Conference on Computer and Communications Security (CCS)},
  year      = {2016}
}

@misc{hazell2023spear,
  title         = {Spear Phishing With Large Language Models},
  author        = {Hazell, Julian},
  year          = {2023},
  eprint        = {2305.06972},
  archivePrefix = {arXiv},
  howpublished  = {arXiv preprint arXiv:2305.06972}
}

@inproceedings{he2024webvoyager,
  title     = {{WebVoyager}: Building an End-to-End Web Agent with Large Multimodal Models},
  author    = {He, Hongliang and Yao, Wenlin and Ma, Kaixin and others},
  booktitle = {Annual Meeting of the Association for Computational Linguistics (ACL)},
  pages     = {6864--6890},
  year      = {2024},
  doi       = {10.18653/v1/2024.acl-long.371}
}

@article{heiding2024phishing,
  title     = {Devising and Detecting Phishing Emails Using Large Language Models},
  author    = {Heiding, Fred and Schneier, Bruce and Vishwanath, Arun and Bernstein, Jeremy and Park, Peter S.},
  journal   = {IEEE Access},
  volume    = {12},
  pages     = {42131--42146},
  year      = {2024},
  doi       = {10.1109/ACCESS.2024.3375882}
}

@article{hyland2012homepages,
  title     = {Individuality or Conformity? Identity in Personal and University Academic Homepages},
  author    = {Hyland, Ken},
  journal   = {Computers and Composition},
  volume    = {29},
  number    = {4},
  pages     = {309--322},
  year      = {2012},
  doi       = {10.1016/j.compcom.2012.10.002}
}

@article{jagatic2007social,
  title     = {Social Phishing},
  author    = {Jagatic, Tom N. and Johnson, Nathaniel A. and Jakobsson, Markus and Menczer, Filippo},
  journal   = {Communications of the ACM},
  volume    = {50},
  number    = {10},
  pages     = {94--100},
  year      = {2007},
  doi       = {10.1145/1290958.1290968}
}

@inproceedings{jueckstock2021realistic,
  title     = {Towards Realistic and Reproducible Web Crawl Measurements},
  author    = {Jueckstock, Jordan and Sarker, Shaown and Snyder, Peter and others},
  booktitle = {Proceedings of the Web Conference (WWW)},
  pages     = {80--91},
  year      = {2021},
  doi       = {10.1145/3442381.3450050}
}

@article{kanich2008spamalytics,
  title     = {Spamalytics: An Empirical Analysis of Spam Marketing Conversion},
  author    = {Kanich, Chris and Kreibich, Christian and Levchenko, Kirill and Enright, Brandon and Voelker, Geoffrey M. and Paxson, Vern and Savage, Stefan},
  journal = {Communications of the ACM},
  pages     = {99--107},
  year      = {2009},
  doi       = {10.1145/1562164.1562190}
}

@inproceedings{kim2025online,
  title     = {When {LLMs} Go Online: The Emerging Threat of Web-Enabled {LLMs}},
  author    = {Kim, Hanna and Song, Minkyoo and Na, Seung Ho and Shin, Seungwon and Lee, Kimin},
  booktitle = {USENIX Security Symposium (USENIX Security)},
  pages     = {1729--1748},
  year      = {2025}
}

@article{klein2014rot,
  title     = {Scholarly Context Not Found: One in Five Articles Suffers from Reference Rot},
  author    = {Klein, Martin and Van de Sompel, Herbert and Sanderson, Robert and others},
  journal   = {PLoS ONE},
  volume    = {9},
  number    = {12},
  pages     = {e115253},
  year      = {2014},
  doi       = {10.1371/journal.pone.0115253}
}

@inproceedings{lauinger2017registrars,
  title     = {Game of Registrars: An Empirical Analysis of Post-Expiration Domain Name Takeovers},
  author    = {Lauinger, Tobias and Chaabane, Abdelberi and Buyukkayhan, Ahmet Salih and Onarlioglu, Kaan and Robertson, William},
  booktitle = {USENIX Security Symposium (USENIX Security)},
  pages     = {865--880},
  year      = {2017}
}

@article{ley2009dblp,
  title     = {{DBLP}: Some Lessons Learned},
  author    = {Ley, Michael},
  journal   = {Proceedings of the VLDB Endowment (PVLDB)},
  volume    = {2},
  number    = {2},
  pages     = {1493--1500},
  year      = {2009},
  doi       = {10.14778/1687553.1687577}
}

@inproceedings{li2021goodbot,
  title     = {Good Bot, Bad Bot: Characterizing Automated Browsing Activity},
  author    = {Li, Xigao and Azad, Babak Amin and Rahmati, Amir and Nikiforakis, Nick},
  booktitle = {IEEE Symposium on Security and Privacy (S\&P)},
  pages     = {1589--1605},
  year      = {2021},
  doi       = {10.1109/SP40001.2021.00079}
}

@inproceedings{liu2016dangling,
  title     = {All Your {DNS} Records Point to Us: Understanding the Security Threats of Dangling {DNS} Records},
  author    = {Liu, Daiping and Hao, Shuai and Wang, Haining},
  booktitle = {ACM SIGSAC Conference on Computer and Communications Security (CCS)},
  pages     = {1414--1425},
  year      = {2016},
  doi       = {10.1145/2976749.2978387}
}

@inproceedings{liu2025somesite,
  title     = {Somesite {I} Used To Crawl: Awareness, Agency and Efficacy in Protecting Content Creators From {AI} Crawlers},
  author    = {Liu, Enze and Luo, Elisa and Shan, Shawn and Voelker, Geoffrey M. and Zhao, Ben Y. and Savage, Stefan},
  booktitle = {ACM Internet Measurement Conference (IMC)},
  pages     = {78--99},
  year      = {2025},
  doi       = {10.1145/3730567.3732913}
}

@inproceedings{longpre2024consent,
  title     = {Consent in Crisis: The Rapid Decline of the {AI} Data Commons},
  author    = {Longpre, Shayne and Mahari, Robert and Lee, Ariel and others},
  booktitle = {Advances in Neural Information Processing Systems (NeurIPS), Datasets and Benchmarks Track},
  year      = {2024}
}

@inproceedings{lukas2023pii,
  title     = {Analyzing Leakage of Personally Identifiable Information in Language Models},
  author    = {Lukas, Nils and Salem, Ahmed and Sim, Robert and Tople, Shruti and Wutschitz, Lukas and Zanella-B{\'e}guelin, Santiago},
  booktitle = {IEEE Symposium on Security and Privacy (S\&P)},
  pages     = {346--363},
  year      = {2023},
  doi       = {10.1109/SP46215.2023.10179300}
}

@incollection{mcfadden1974logit,
  title     = {Conditional Logit Analysis of Qualitative Choice Behavior},
  author    = {McFadden, Daniel},
  booktitle = {Frontiers in Econometrics},
  editor    = {Zarembka, Paul},
  publisher = {Academic Press},
  address   = {New York},
  pages     = {105--142},
  year      = {1974}
}

@misc{muehlemann2008nine,
  title        = {Nine Ways to Obfuscate E-Mail Addresses Compared},
  author       = {M{\"u}hlemann, Silvan},
  howpublished = {Blog post, \url{https://blog.muehlemann.com/2008/07/20/ten-methods-to-obfuscate-e-mail-addresses-compared/}},
  year         = {2008}
}

@inproceedings{polakis2010harvest,
  title     = {Using Social Networks to Harvest Email Addresses},
  author    = {Polakis, Iasonas and Kontaxis, Georgios and Antonatos, Spiros and Gessiou, Eleni and Petsas, Thanasis and Markatos, Evangelos P.},
  booktitle = {ACM Workshop on Privacy in the Electronic Society (WPES)},
  pages     = {11--20},
  year      = {2010},
  doi       = {10.1145/1866919.1866922}
}

@incollection{poornachandran2016obfuscation,
  title     = {Effectiveness of Email Address Obfuscation on Internet},
  author    = {Poornachandran, Prabaharan and Raj, Deepak and Pal, Soumajit and Ashok, Aravind},
  booktitle = {Innovations in Computer Science and Engineering},
  series    = {Advances in Intelligent Systems and Computing},
  year      = {2016}
}

@inproceedings{prince2005spammers,
  title     = {Understanding How Spammers Steal Your E-Mail Address: An Analysis of the First Six Months of Data from {Project Honey Pot}},
  author    = {Prince, Matthew B. and Dahl, Benjamin M. and Holloway, Lee and Keller, Arthur M. and Langheinrich, Eric},
  booktitle = {Conference on Email and Anti-Spam (CEAS)},
  year      = {2005}
}

@misc{psl2026,
  title        = {Public Suffix List},
  author       = {{Mozilla Foundation}},
  howpublished = {\url{https://publicsuffix.org/}},
  year         = {2026},
  note         = {Accessed October 2026}
}

@article{rousseeuw1987silhouettes,
  title     = {Silhouettes: A Graphical Aid to the Interpretation and Validation of Cluster Analysis},
  author    = {Rousseeuw, Peter J.},
  journal   = {Journal of Computational and Applied Mathematics},
  volume    = {20},
  pages     = {53--65},
  year      = {1987},
  doi       = {10.1016/0377-0427(87)90125-7}
}

@inproceedings{searles2023captcha,
  title     = {An Empirical Study \& Evaluation of Modern {CAPTCHAs}},
  author    = {Searles, Andrew and Nakatsuka, Yoshimichi and Ozturk, Ercan and Paverd, Andrew and Tsudik, Gene and Enkoji, Ai},
  booktitle = {USENIX Security Symposium (USENIX Security)},
  year      = {2023}
}

@article{shrout1979icc,
  title     = {Intraclass Correlations: Uses in Assessing Rater Reliability},
  author    = {Shrout, Patrick E. and Fleiss, Joseph L.},
  journal   = {Psychological Bulletin},
  volume    = {86},
  number    = {2},
  pages     = {420--428},
  year      = {1979},
  doi       = {10.1037/0033-2909.86.2.420}
}

@inproceedings{shue2009spamology,
  title     = {Spamology: A Study of Spam Origins},
  author    = {Shue, Craig A. and Gupta, Minaxi and Lubia, John T. and Kong, C. H. and Y\"{u}ksel, As{\i}m Sinan},
  booktitle = {Conference on Email and Anti-Spam (CEAS)},
  year      = {2009}
}

@inproceedings{smith2007tesseract,
  title     = {An Overview of the {Tesseract} {OCR} Engine},
  author    = {Smith, Ray},
  booktitle = {International Conference on Document Analysis and Recognition (ICDAR)},
  volume    = {2},
  pages     = {629--633},
  year      = {2007},
  doi       = {10.1109/ICDAR.2007.4376991}
}

@inproceedings{staab2024memorization,
  title     = {Beyond Memorization: Violating Privacy via Inference with Large Language Models},
  author    = {Staab, Robin and Vero, Mark and Balunovi{\'c}, Mislav and Vechev, Martin},
  booktitle = {International Conference on Learning Representations (ICLR)},
  year      = {2024}
}

@inproceedings{steinacker2026robots,
  title     = {Is Misinformation More Open? A Study of robots.txt Gatekeeping on the Web},
  author    = {Steinacker-Olsztyn, Nicolas and Gosain, Devashish and Dao, Ha},
  booktitle = {Proceedings of the ACM Web Conference (WWW)},
  year      = {2026}
}

@inproceedings{tang2008arnetminer,
  title     = {{ArnetMiner}: Extraction and Mining of Academic Social Networks},
  author    = {Tang, Jie and Zhang, Jing and Yao, Limin and Li, Juanzi and Zhang, Li and Su, Zhong},
  booktitle = {ACM SIGKDD International Conference on Knowledge Discovery and Data Mining (KDD)},
  pages     = {990--998},
  year      = {2008},
  doi       = {10.1145/1401890.1402008}
}

@inproceedings{vastel2020fpcrawlers,
  title     = {{FP-Crawlers}: Studying the Resilience of Browser Fingerprinting to Block Crawlers},
  author    = {Vastel, Antoine and Rudametkin, Walter and Rouvoy, Romain and Blanc, Xavier},
  booktitle = {Workshop on Measurements, Attacks, and Defenses for the Web (MADWeb)},
  year      = {2020},
  doi       = {10.14722/madweb.2020.23010}
}

@inproceedings{vonahn2003captcha,
  title     = {{CAPTCHA}: Using Hard {AI} Problems for Security},
  author    = {von Ahn, Luis and Blum, Manuel and Hopper, Nicholas J. and Langford, John},
  booktitle = {Advances in Cryptology --- EUROCRYPT 2003},
  series    = {LNCS},
  volume    = {2656},
  pages     = {294--311},
  year      = {2003},
  doi       = {10.1007/3-540-39200-9_18}
}

@article{vonahn2008recaptcha,
  title     = {{reCAPTCHA}: Human-Based Character Recognition via Web Security Measures},
  author    = {von Ahn, Luis and Maurer, Benjamin and McMillen, Colin and Abraham, David and Blum, Manuel},
  journal   = {Science},
  volume    = {321},
  number    = {5895},
  pages     = {1465--1468},
  year      = {2008},
  doi       = {10.1126/science.1160379}
}

@article{ward1963hierarchical,
  title     = {Hierarchical Grouping to Optimize an Objective Function},
  author    = {Ward, Joe H., Jr.},
  journal   = {Journal of the American Statistical Association (JASA)},
  volume    = {58},
  number    = {301},
  pages     = {236--244},
  year      = {1963},
  doi       = {10.1080/01621459.1963.10500845}
}

@inproceedings{zeber2020representativeness,
  title     = {The Representativeness of Automated Web Crawls as a Surrogate for Human Browsing},
  author    = {Zeber, David and Bird, Sarah and Oliveira, Camila and others},
  booktitle = {Proceedings of The Web Conference (WWW)},
  pages     = {167--178},
  year      = {2020},
  doi       = {10.1145/3366423.3380104}
}

@inproceedings{zheng2024seeact,
  title     = {{GPT-4V(ision)} is a Generalist Web Agent, if Grounded},
  author    = {Zheng, Boyuan and Gou, Boyu and Kil, Jihyung and Sun, Huan and Su, Yu},
  booktitle = {International Conference on Machine Learning (ICML)},
  year      = {2024}
}

@inproceedings{zhou2024webarena,
  title     = {{WebArena}: A Realistic Web Environment for Building Autonomous Agents},
  author    = {Zhou, Shuyan and Xu, Frank F. and Zhu, Hao and others},
  booktitle = {International Conference on Learning Representations (ICLR)},
  year      = {2024}
}

\appendix
\setcounter{dbltopnumber}{4}\renewcommand{\dbltopfraction}{0.98}\renewcommand{\dblfloatpagefraction}{0.9}\renewcommand{\textfraction}{0.05}
\section{Details of the Taxonomy and the Pipeline}
\label{app:details}

\begin{figure*}[t]
\centering
\includegraphics[width=\textwidth]{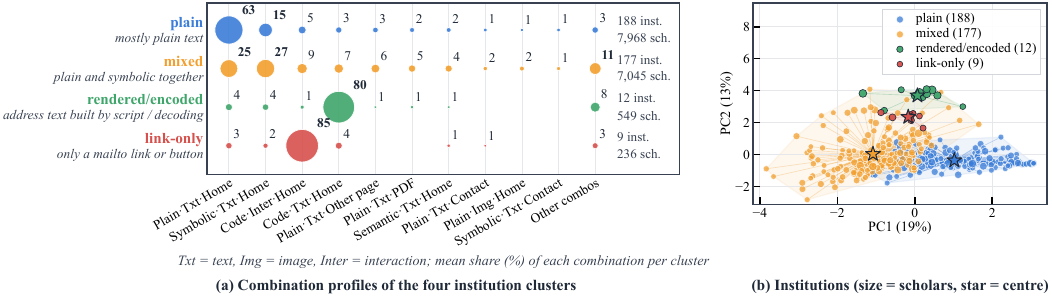}
\Description{Bubble chart of combination shares for four institution clusters and a scatter plot of institutions on two principal components.}
\caption{Left: mean within-institution share of the ten globally most common content--form--location combinations plus an ``other'' category, for each cluster. Right: the first two principal components of standardized features (used for this projection only), colored by the clusters obtained from unstandardized proportions.}
\label{fig:ic3}
\end{figure*}

\begin{table*}[t]
\centering
\caption{The three audit rounds. Each round draws 500 pages with a new seed from pages that no earlier round and no earlier development step had used, and the auditor judges the labels as the pipeline held them at that time. The last column is the share of correct \cSem addresses whose content label the auditor accepts.}
\label{tab:rounds}
\small
\begin{tabular*}{\textwidth}{@{\extracolsep{\fill}}llllll@{}}
\toprule
\textbf{Round} & \textbf{Address correct [95\% CI]} & \textbf{Content} & \textbf{Form} & \textbf{Location} & \textbf{\cSem content} \\
\midrule
1 (before the fixes below) & 92.2\% \ci{2.3}{92.2}{88.4}{94.8}{80}{100}{cplain} {\scriptsize[88.4, 94.8]}\enspace & \pbar{1.05}{0.949}{ccode}\,94.9\% & \pbar{1.05}{0.970}{ccode}\,97.0\% & \pbar{1.05}{0.975}{ccode}\,97.5\% & \pbar{1.05}{0.848}{csem}\,85\% \\
2 (after cleanup and location fixes) & 91.1\% \ci{2.3}{91.1}{87.0}{94.0}{80}{100}{cplain} {\scriptsize[87.0, 94.0]}\enspace & \pbar{1.05}{0.958}{ccode}\,95.8\% & \pbar{1.05}{0.968}{ccode}\,96.8\% & \pbar{1.05}{0.961}{ccode}\,96.1\% & \pbar{1.05}{0.764}{csem}\,76\% \\
3 (frozen system, final) & 91.6\% \ci{2.3}{91.6}{87.4}{94.5}{80}{100}{cplain} {\scriptsize[87.4, 94.5]}\enspace & \pbar{1.05}{0.962}{ccode}\,96.2\% & \pbar{1.05}{0.981}{ccode}\,98.1\% & \pbar{1.05}{0.982}{ccode}\,98.2\% & \pbar{1.05}{0.882}{csem}\,88\% \\
\bottomrule
\end{tabular*}
\end{table*}

\begin{figure*}[t]
\centering
\includegraphics[width=\textwidth]{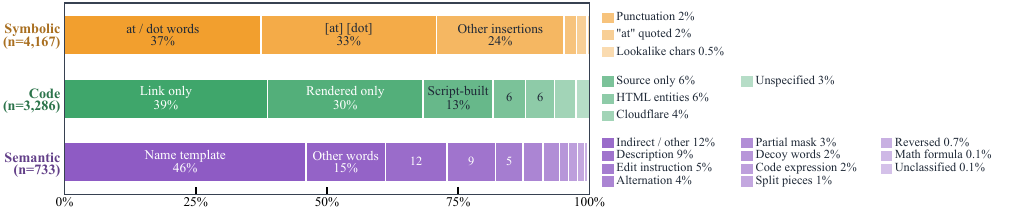}
\Description{Stacked bars of subtype shares within the Symbolic, Code, and Semantic classes.}
\caption{Subtype distributions among scholars in each non-plain content class; each row uses that class as its denominator.}
\label{fig:sub}
\end{figure*}

\subsection{Decision Rules}
\label{app:rules}

Beyond the definitions in Section~\ref{sec:dims} and Algorithm~\ref{alg:priority}, we apply the following rules to every recovered address, whatever layer recovered it.
The substitution lexicon contains \emph{at}, \emph{dot}, \emph{period}, \emph{point}, the misspelling \emph{doc}, brackets, quotes, look-alike characters, icon-font at signs, and inserted separators; presentations that use only these are \cSubst, whereas other substitutions, templates, masks, reconstruction instructions, inserted words, alternatives, and expressions displayed for a reader are \cSem, even when a rule reconstructs them.
Source-only strings in comments, attributes, hidden elements, metadata, or scripts are \cCode with form \fAct, and a dynamically inserted address has form \fText when directly visible and \fAct when a click is required.
When the served HTML is a bot challenge, identified by the markers of common bot-management products, or a redirect stub, we classify the page that a browser obtained and record the barrier (\S\ref{sec:rq4}).

\begin{algorithm}[t]
\caption{Classifying the email addresses of one scholar}
\algrenewcommand{\algorithmiccomment}[1]{\hfill\makebox[2.95cm][l]{$\triangleright$ #1}}
\label{alg:priority}
\small
\begin{algorithmic}[1]
\Require own addresses $E$; presentations $P(e)$ of each $e$; page and techniques $T(p)$ of each $p$ \Statex $C$: \cPlain$<$\cSubst$<$\cCode$<$\cSem \Statex $F$: \fText$<$\fImage$<$\fAct \Statex $L$: Homepage$<$Contact$<$CV$<$Same-site$<$External$<$PDF \For{each $e \in E$} \For{each $p \in P(e)$} \Comment{case 1: hardest technique} \State $c(p) \gets \max_C \{c(t) : t \in T(p)\}$ \State $f(p) \gets \max_F \{f(t) : t \in T(p)\}$ \State $l(p) \gets L(\mathrm{page}(p))$ \EndFor \State $l^* \gets \min_L \{l(p) : p \in P(e)\}$ \Comment{case 2: easiest location} \State $p^* \gets \arg\min_{p:\, l(p)=l^*} \big(c(p), f(p)\big)$ \Comment{then content, form} \State $\mathit{label}(e) \gets (c(p^*), f(p^*), l^*)$ \EndFor \State \Return $\arg\min_{e \in E} \big(l^*(e), c(p^*_e), f(p^*_e)\big)$ \Comment{case 3: easiest address}
\end{algorithmic}
\end{algorithm}

\subsection{Extraction Layers in Detail}
\label{app:extract}

A full rule-layer run processes the \nTotal records in about \ruleMinutes minutes on four cores, whereas the LLM layer processes about \llmRate pages per minute on its (different) workload.

\noindent\textbf{L1.}
Rules match JavaScript assembly patterns such as character arrays, string concatenation, helper calls, split user-name and domain attributes, and encoded \texttt{mailto} values; for \cSem, they cover only name-filled templates and partial masks.
Ownership uses matches between the local part and the scholar's name, nearby role terms, generic mailbox names, and the number of scholars' pages containing the address; a lone non-generic address on a personal page is accepted with low confidence (\pConfLow) and checked by hand.

\noindent\textbf{L2.}
L2 handles images, descriptions or templates outside the rules, addresses available only through interactions or linked pages, and ambiguous ownership on pages listing several people.
The models receive the homepage text, the HTML around email clues, up to four linked pages, and the text that a vision model extracts from images next to an email label; when an image can no longer be fetched, we apply optical character recognition to the archived screenshot.
Both models share one prompt that contains the taxonomy and requests JSON output with the scholar's addresses and one label per dimension; page text is placed inside delimiters and treated as data, and the models return no address when the evidence is insufficient.
Cases without a majority after the Opus assessment are marked as low confidence; the vision models are Claude Sonnet 5.5 with GLM-5V-Turbo and GPT-5.6 Sol as fallbacks.

\noindent\textbf{L3.}
On department or university sites, the agent may read people lists and run at most two site searches (\rescueInstPages-page budget); on personal sites, it prioritizes contact and CV pages, then linked personal websites and PDFs (\rescuePersPages pages).
Each record has a \rescueSeconds-second limit, navigation is restricted to URLs found in retrieved pages, and public ORCID records are the only external index; GLM-5.3 checks the retrieved evidence.
A language model assigned the \cSem subtypes, and a second model agreed on \semCheckRate of \semCheckN sampled addresses.

\begin{table}[h!]
\centering
\caption{Content classes recovered by each extraction layer.}
\label{tab:stage}
\small
\begin{tabular*}{\columnwidth}{@{\extracolsep{\fill}}lrcccc@{}}
\toprule
\textbf{Layer} & \textbf{Scholars} & \textbf{\cPlain} & \textbf{\cSubst} & \textbf{\cCode} & \textbf{\cSem} \\
\midrule
L1 rules & \nSrcRule & 56.2\% & 24.6\% & 18.6\% & 0.7\% \\
L2 LLMs & \nSrcLlm & 39.9\% & 24.8\% & 18.0\% & 17.3\% \\
L3 agent & \nSrcRescue & 56.1\% & 13.5\% & 19.3\% & 11.1\% \\
\bottomrule
\end{tabular*}
\end{table}

\begin{figure*}[t]
\centering
\includegraphics[width=\textwidth]{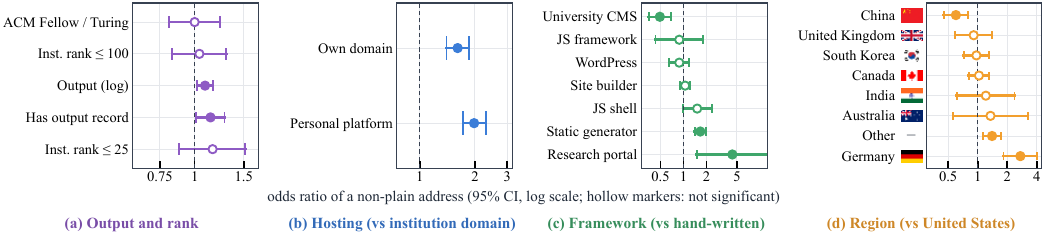}
\Description{Forest plots of odds ratios for a non-plain address across output and rank, hosting, framework, and region predictors.}
\caption{Odds ratios for a non-plain selected presentation from one logistic model, by predictor group (95\% intervals clustered by institution; hollow markers are not significant).}
\label{fig:forest}
\end{figure*}

\begin{figure*}[t]
\centering
\includegraphics[width=\textwidth]{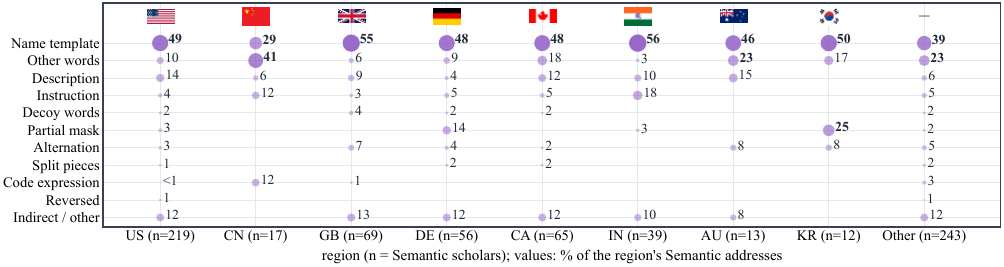}
\Description{Bubble matrix of Semantic subtype shares across regions, with flags above each region.}
\caption{Share of \cSem scholars assigned each subtype within a region; $n$ counts the region's \cSem scholars.}
\label{fig:heat}
\end{figure*}

\subsection{Dataset Details}
\label{app:data}

Spelling variants of one name yield repeated listings (\pDupSame of the records); of the \nSharedUnique scholars with a shared URL, \nDupShared share it with a different surname and \nDupSurname with the same surname.
Among unreachable hosts, we record \nUnDns DNS failures, \nUnTimeout timeouts, \nUnTls certificate failures, and \nUnOther other network errors; other failures are \texttt{robots.txt} exclusions (\nSRobotsDisallowed), client errors (\nSHttpError), server errors (\nSServerError), and non-HTML documents (\nSNonHtml).
Before the final step, 1,954 reachable scholars had no address; a single-model search of \nAsevenPool of them (registered page and its links only; the 21 human-sample records excluded) recovered evidence-supported addresses for \nAsevenFound.
Reachable homepages cover \nInstAcc of the \nInstAll affiliations.

\subsection{Statistical Estimation}
\label{app:estimation}

Each address-bearing scholar contributes one label triple to the distribution shares.
For validation, we sample the four content classes and the no-address group with equal allocation.
For stratum $h$ with frame size $N_h$, $n_h$ evaluable cases, and $s_h$ agreements, we compute $w_h=N_h/\sum_j N_j$, $\hat p_h=s_h/n_h$, and $\hat p=\sum_h w_h\hat p_h$, renormalizing the weights over strata with evaluable labels for conditional label estimates.
We compute $\widehat{\mathrm{Var}}(\hat p)=\sum_h w_h^2\hat p_h(1-\hat p_h)/n_h$ and $n_{\mathrm{eff}}=\hat p(1-\hat p)/\widehat{\mathrm{Var}}(\hat p)$ and use them in the Wilson interval with $z=1.96$; when the variance is zero, we use the number of evaluable cases.
Label agreement in the double-annotation subset is an unweighted proportion among the \hvContentN evaluable cases.

\subsection{Institution Clusters}
\label{app:clusters}

For the \nInstFifteen institutions with at least 15 address-bearing scholars, we represent each institution by its shares of the ten most frequent content--form--location combinations in the corpus plus one residual share.
We apply Ward hierarchical clustering~\cite{ward1963hierarchical} with Euclidean distance to these unstandardized proportions and select $k=\icKclusters$, the cut among $k=3,\ldots,8$ that maximizes the mean silhouette score~\cite{rousseeuw1987silhouettes} (\icSil); the names describe the resulting clusters (Figure~\ref{fig:ic3}).
A \emph{plain} cluster of \icPlainInst institutions has \icPlainPlain \cPlain, and a \emph{mixed} cluster of \icMixInst institutions (\icMixSch scholars) has \icMixPlain \cPlain and \icMixSym \cSubst.
Two small clusters are dominated by \cCode but differ in form.
In the \emph{rendered/encoded} cluster (\icRenInst institutions), the browser exposes the address as readable text after rendering or decoding (mean Code--Text--Homepage share \icRenTop); in the \emph{link-only} cluster (\icMailInst institutions), a visitor must follow an ``Email'' link or control (Code--Interaction--Homepage \icMailTop).
The largest host holds on average \icRenConc and \icMailConc of these clusters' scholars, consistent with shared profile templates.

\subsection{Additional Results}
\label{app:more}

Table~\ref{tab:rounds} reports each audit round; \cSem content, the hardest label, is accepted for 76--88\% of correct \cSem addresses.
Figure~\ref{fig:sub} lists every non-plain subtype, extending the outer ring of Figure~\ref{fig:dist}a.
Figure~\ref{fig:forest} gives the odds ratios of the full logistic model in Section~\ref{sec:rq3}: portals, static generators, and personal hosting are associated with higher odds, university CMSs with lower odds, and output and rank have odds ratios near one.
Figure~\ref{fig:heat} breaks the \cSem subtypes down by region: name templates dominate everywhere, but other words reach 41\% in China and partial masks 25\% in South Korea, while edit instructions peak in India (18\%) and descriptions in Australia (15\%).

\subsection{Ethics and Restricted Artifacts}
\label{app:ethics}
\label{app:reproducibility}

An early configuration rendered \nRobotsRendered of the \nSRobotsDisallowed homepages disallowed by \texttt{robots.txt}; all are excluded from the results.
Ordinary browser execution sometimes passed JavaScript challenges automatically, but we did not solve CAPTCHAs, authenticate, or use other mechanisms to bypass access controls.
The model-assisted stages sent public page material to the named providers, which remains a privacy consideration.
We do not release the operational code, prompts, page archives, extracted addresses, or individual-level records, because a consolidated corpus and a ready-to-run extractor would ease bulk harvesting and expose scholars' personal data beyond the context in which they chose to publish it; the documented methods suffice for reimplementation.

\end{document}